\documentclass[nonacm]{acmart}

\pdfoutput=1
\AtBeginDocument{%
  }

\setcopyright{acmcopyright}
\copyrightyear{2022}
\acmYear{2022}
\acmDOI{XXXXXXX.XXXXXXX}

\usepackage{color}
\usepackage{appendix}
\usepackage{tabularx}
\usepackage{multirow}
\usepackage{xspace}
\usepackage{colortbl}
\usepackage{subcaption}
\usepackage{enumitem}
\usepackage{hhline}

\newcommand{\cmark}{\checkmark}
\newcommand{\xmark}{\textsf{X}}
\newcommand{\bF}[1]{\xmark$^{#1}$}
\newcommand{\rF}[1]{\textcolor{red}{\xmark}$^{#1}$}
\newcommand{\bT}[1]{\cmark$^{#1}$}

\newcommand{\rO}[1]{\textsf{\textcolor{red}{O}$^{#1}$}}
\newcommand{\bO}[1]{\textsf{O$^{#1}$}}
\newcommand{\SMDP}{\text{SM-DP+}\xspace}
\newcommand{\sign}[2]{\ensuremath{\text{Sign}(#2; #1)}}
\newcommand{\enc}[2]{\ensuremath{\text{E}(#2; #1)}}
\newcommand{\mac}[2]{\ensuremath{\text{MAC}(#2; #1)}}

\newcommand{\cert}[1]{\ensuremath{\mathit{Cert}_{\mathit{#1}}}}
\newcommand{\Sa}{\ensuremath{\mathit{Sa}}\xspace}
\newcommand{\Sp}{\ensuremath{\mathit{Sp}}\xspace}

\newcommand{\CI}{\ensuremath{\mathit{CI}}\xspace}
\newcommand{\SKI}[1]{\ensuremath{\mathit{SKI}_{\mathit{#1}}}}
\newcommand{\SK}[1]{\ensuremath{\mathit{SK\!}_{\mathit{#1}}}}

\newcommand{\UserId}{\ensuremath{\mathit{userId}}\xspace}
\newcommand{\Iac}{\ensuremath{\mathit{I}_{\mathit{ac}}}\xspace}

\newcommand{\Seq}{\ensuremath{\mathit{Seq}}\xspace}
\newcommand{\oid}{\ensuremath{\mathit{serverOID}}\xspace}
\newcommand{\mnoid}{\ensuremath{\mathit{mnoId}}\xspace}

\newcommand{\attacker}{\ensuremath{\mathit{attacker}}\xspace}
\newcommand{\event}[2]{\ensuremath{\mathit{event(\text{#1}(#2))}}\xspace}
\newcommand{\injevent}[2]{\ensuremath{\mathit{inj\mbox{-}event(\text{#1}(#2))}}\xspace}
\newcommand{\freevar}[1]{\ensuremath{\mathit{\mbox{\_}}}\xspace}
\newcommand{\AND}{\;\;\text{\&\&}\;\;}

\newcommand{\IMPLY}{\;\Rightarrow\;}
\newcommand{\false}{\text{false}}
\newcommand{\resultstableref}[1]{($^{#1}$~in \hyperlink{target:results_tables}{results})}
\newcommand{\quotebox}[2]{\begin{center}\vspace{0.1\baselineskip}\begin{minipage}{0.90\textwidth}%
\noindent``\textit{#1}'' #2\end{minipage}\end{center}}

\newcommand{\rr}[1]{#1} 
\newcommand{\recommend}[1]{\textbf{#1}}

\newcommand{\red}[1]{\textcolor{black}{#1}}

\newcommand{\Rec}[1]{\textcolor{violet}{#1}}
\newcommand{\mycomment}[3]{\textcolor{#1}{#2:~#3}}
\newcommand{\ta}[1]{\mycomment{red}{TA}{#1}}

\newcommand{\hidetext}[1]{#1}
\renewcommand{\hidetext}[1]{}

\begin{document}

\title{Security Analysis of the Consumer Remote SIM Provisioning Protocol}


\author{Abu Shohel Ahmed}
\authornote{Corresponding Author}
\email{abu.ahmed@aalto.fi}
\affiliation{%
  \institution{Aalto University}
  \city{Espoo}
  \country{Finland}
}



\author{Aleksi Peltonen}
\email{aleksi.peltonen@aalto.fi}
\affiliation{%
  \institution{Aalto University}
  \city{Espoo}
  \country{Finland}
}

\author{Mohit Sethi}
\email{mohit.sethi@aalto.fi}
\affiliation{%
  \institution{Aalto University}
  \city{Espoo}
  \country{Finland}
}

\author{Tuomas Aura}
\email{tuomas.aura@aalto.fi}
\affiliation{%
  \institution{Aalto University}
  \city{Espoo}
  \country{Finland}
}


\begin{abstract}
Remote SIM provisioning (RSP) for consumer devices is the protocol specified by the GSM Association for downloading SIM profiles into a secure element in a mobile device. The process is commonly known as eSIM, and it is expected to replace removable SIM cards. The security of the protocol is critical because the profile includes the credentials with which the mobile device will authenticate to the mobile network. In this paper, we present a formal security analysis of the consumer RSP protocol. We model the multi-party protocol in applied pi calculus, define formal security goals, and verify them in ProVerif. The analysis shows that the consumer RSP protocol protects against a network adversary when all the intended participants are honest. However, we also model the protocol in realistic partial compromise scenarios where the adversary controls a legitimate participant or communication channel. The security failures in the partial compromise scenarios reveal weaknesses in the protocol design. The most important observation is that the security of RSP depends unnecessarily on it being encapsulated in a TLS tunnel. Also, the lack of pre-established identifiers means that a compromised download server anywhere in the world or a compromised secure element can be used for attacks against RSP between honest participants. Additionally, the lack of reliable methods for verifying user intent can lead to serious security failures. Based on the findings, we recommend practical improvements to RSP implementations, future versions of the specification, and mobile operator processes to increase the robustness of eSIM security.

\end{abstract}

\ccsdesc[500]{Security and privacy~Formal security models}
\ccsdesc[500]{Security and privacy~Mobile and wireless security}
\ccsdesc[300]{Security and privacy~Security protocols}
\ccsdesc[300]{Security and privacy~Authentication}

\keywords{mobile communication, eSIM, formal modeling and verification}

\maketitle


\section{Introduction}
\label{sec:Introduction}

Subscriber identity module (SIM) is the secure element in a mobile phone or device that contains the identifiers and cryptographic credentials with which the \textit{user equipment} (UE) authenticates the mobile subscriber to the \textit{mobile network operator} (MNO). For a long time, the SIM has been a miniature smart card that is inserted into the mobile device, and by changing the SIM, it has been possible to use the same device for different mobile subscribers and networks. However, in new mobile devices, the removable SIM is being replaced by the \textit{embedded SIM} (eSIM). It is a secure element integrated into the circuit board of the mobile device, and it can be programmed with \textit{SIM profiles} that contain the identifiers and credentials.

\textit{Remote SIM provisioning} (RSP)~\cite{esim-whitepaper} is the protocol specified by the GSM Association (GSMA) for the purpose of downloading and installing SIM profiles into the secure element. Remote SIM provisioning reduces the logistical and production costs, and it gives subscribers the flexibility of changing operators online~\cite{vesselkov2015value}. There are two variants of the RSP protocol: the machine-to-machine (M2M) version~\cite{RSP-01,RSP-02} is used in remotely controllable devices, and the consumer version~\cite{RSP-21v2,RSP-22v2} is used in consumer devices such as smartphones, tablets, and wearables. These two protocol variants differ in how much participation is expected from the end users. The M2M variant does not need interaction with the end user for SIM profile management. In contrast, the consumer RSP protocol requires the end user to trigger the management operations such as remote provisioning, enabling, disabling, and deletion of SIM profiles on the mobile device. 

The SIM profile contains security-critical information such as the international mobile subscriber identity (IMSI) and subscriber key $K_i$, which is a shared secret between the UE and the MNO. The UE uses these credentials from the SIM profile in the authentication and key agreement (AKA) procedure to enable the subscriber's access to the mobile network~\cite{3GPP-TS-33401}. Thus, the secure delivery of the SIM profile to the mobile device is of the utmost importance. Unwanted exposure or tampering of the SIM profile or the credentials within it could lead to identity theft, billing fraud, eavesdropping, and various privacy violations against the mobile subscriber. In one documented security failure~\cite{Heist2015}, the attackers captured credentials of SIM profiles by penetrating the internal networks of a SIM manufacturer. In another case~\cite{meyer2018attacks}, the attackers were able to carry out memory-exhaustion attacks against the secure element using flaws in the M2M RSP protocol. These attacks highlight the importance of careful design and analysis for the RSP protocols.

This paper investigates the security of the \textit{consumer RSP protocol}. We develop a formal protocol model and requirements using applied pi calculus and the ProVerif verification tool~\cite{blanchet2005proverif} for security-protocol modeling and verification. Our formal verification shows that the consumer RSP protocol meets the expected security goals in the presence of a network attacker. However, we also analyze the protocol under partial compromise scenarios where one of the system components is not completely trustworthy. This analysis provides insights into the protocol design and also reveals several weaknesses that could lead to security failures in realistic scenarios.
Specifically, our contributions are the following:
\begin{enumerate}
  \item We extract the core message flows and content from the consumer RSP protocol specification~\cite{RSP-22v2,RSP-21v2}. Based on this information, we develop a comprehensive formal model that covers the key protocol variants. To our knowledge, this is the first formal model of the consumer RSP protocol.
  \item From the specification, we extract both explicit and implicit security requirements that the RSP protocol must fulfill and use them to define formal security goals. Furthermore, we define partial compromise scenarios that will provide interesting new information about the weaknesses of the protocol.
  \item We present the verification results, which include several unexpected security failures under the partial compromise scenarios. When a formal security goal fails, we analyze and explain the attacks and their root causes in a way that is easy to relate to the protocol specification.
  \item Finally, we provide practical recommendations for future development of the RSP specification, as well as, implementation guidance to increase its robustness against attacks. The defensive effect of these recommendations is verified in the formal model.
\end{enumerate}

The rest of the article is organized as follows. Section~\ref{sec:background} provides background information on protocol modeling, ProVerif, and SIM provisioning security. Section~\ref{sec:protocol} explains the RSP architecture and message flows and the different ways in which the user can initiate profile download. Section~\ref{sec:formalmodel} describes the formal model and the modeled partial compromise scenarios, and Section~\ref{sec:formalgoals} defines the formal security goals. Section~\ref{sec:results} presents the verification results, as well as, recommendations for improving the specification and its secure implementation. Section~\ref{sec:discussion} discusses the implications of the results and other aspects of the modeling process. Finally, Section~\ref{sec:conclusion} concludes the article.


\section{Background}
\label{sec:background}

This section provides an overview of related work on protocol verification, the formal modeling and verification tool ProVerif, and previous research on SIM provisioning security.

\subsection{Related work on security protocol analysis}
\label{sec:relatedWork}

Security protocols are prone to subtle design errors. The academic research in this area started from the publication of the Needham and Schroeder protocol~\cite{needham1978using} in 1978. Since then, researchers have discovered protocol flaws as well as proposed improved designs and security requirements \cite{denning1981timestamps,woo1992authentication,abadi1996prudent,lowe1996breaking,woo1997authentication,blake1999unknown,krawczyk2003sigma}. Nevertheless, logical flaws continue to appear in security protocols, at least in the early stages of their development. These flaws may be simple mistakes, but sometimes they reflect our improved understanding of the security assumptions and requirements for a specific type of protocol. Because of this history, it is generally agreed that security protocols should undergo thorough scrutiny by the security research community.

One approach to quality assurance is formal analysis~\cite{burrows1990logic,meadows1994formal,mao_modern_2003,ling_cryptographic_2012}, which can verify that the protocol meets well-defined security goals or uncover weaknesses. A common approach to formal analysis is symbolic model checking~\cite{basin_model_2018}. Model checking tools such as ProVerif~\cite{blanchet2005proverif}, Tamarin~\cite{meier2013tamarin,schmidt2012automated}, and DeepSec~\cite{cheval2018deepsec} perform automated verification of protocol models against formalized security goals.
For example, Bhargavan et al.~\cite{bhargavan2017verified} used ProVerif to verify the Transport Layer Security (TLS) 1.3~\cite{rescorla2018transport} standard candidates, leading to modifications in the protocol. Similarly, Cremers et al.~\cite{cremers2016automated} used the Tamarin model checker and found possible attacks on certificate-based and pre-shared key authentication in TLS. They also modeled the full protocol~\cite{cremers2017comprehensive}. Basin et al.~\cite{basin2018formal,basin2021emv,basin2021card} formalized and verified the security goals for the fifth-generation (5G) mobile network authentication and key agreement (AKA)~\cite{3GPP-TS-33501} protocol and the EMV payment card protocol. Their work also supported standardization by formalizing the security requirements and by proposing improvements. These examples show the benefits of formal modeling and verification. However, written protocol specifications are not always easy to convert into precise formal models, and the adversary model and security goals for the protocols are not always fully understood. Moreover, the number of protocol variants that need to be modeled can be high; for example, Cremers~\cite{cremers2011key} analyzed over forty variants of the Internet key exchange (IKE).

\subsection{Applied pi calculus and the ProVerif tool}
\label{sec:proverif}

ProVerif~\cite{blanchet2005proverif} is a tool for modeling and automatic verification of cryptographic protocols. The protocol participants are modeled as communicating processes in applied pi calculus \cite{abadi2018applied}. The security goals are formalized as queries that will be evaluated against the protocol model. In particular, authentication can be formalized as a correspondence between local events in the concurrent processes.
ProVerif has built-in support for the network adversary, i.e., the Dolev-Yao type adversary~\cite{dolev1983security}, which is in full control of the network and can read, modify, delete, and inject messages and perform cryptographic computations on them. When the verification of a security goal fails, the tool tries to construct an attack trace that shows how the adversary can violate it.
The models are symbolic, and they treat cryptographic primitives, such as signatures and encryption, as abstract functions that are assumed to be secure. This can be contrasted with computational models~\cite{blanchet2007computationally}, which consider the probability of breaking cryptographic primitives. The symbolic approach is good for finding logical flaws in the protocol design.

\subsection{SIM provisioning security}
\label{sec:background_SIM_provisioning}

In standard terminology, the older SIM is a universal integrated circuit card (UICC), and the embedded version is an \textit{embedded UICC} (eUICC). Traditionally, a SIM profile is provisioned to the UICC in the smart-card factory. One provisioned UICC is inserted into the user equipment (UE) for each mobile network subscription. This approach requires physical delivery of the SIM card, and it also limits the user's flexibility in changing subscriptions. An earlier attempt to overcome these shortcomings was the virtualized \textit{soft SIM}~\cite{sim-reprogrammable}. It provides similar functionality to the removable SIM card but without hardware-based secure credential storage. The soft-SIM approach was not adopted widely due to its perceived lack of security~\cite{sim-evolution}. RSP aims to provide the same level of security as the removable SIM card by storing the SIM profile in the eUICC, which is an embedded secure element in the mobile device.

So far, security analysis of RSP has focused on the eUICC component and the M2M variant of the protocol. Vahidian~\cite{vahidian2013evolution} analyzed the security of the eUICC and soft-SIM. \red{They focused on securing the SIM profile within a secure container and do not analyze the protocol security.} The informal threat analysis by GSMA surveys security risks for the M2M RSP~\cite{RSP-01}. \red{The report enumerates potential security failures for the different system components and explains the likely consequences. This work inspired some of the security goals and scenarios in our formal model.} Meyer et al.~\cite{meyer2018attacks} discovered weaknesses in the M2M RSP~\cite{RSP-02} related to denial of service by exhausting the capacity of the eUICC. \red{While there is no overlap with our work, the results motivated us to investigate the security of the consumer RSP version.} Ding et al.~\cite{ding2021formal} model M2M RSP with SPIN~\cite{holzmann1997model}, which, however, is a general-purpose model checker and not intended for verifying security properties~\cite{ninet2019model}. Park et al.~\cite{park2013secure} proposed an alternative to the GlobalPlatform secure channel protocol (SCP)~\cite{platform2006global,SCP11}, which is also used in RSP, with potential scalability and performance gains.


\section{Consumer remote SIM provisioning protocol}
\label{sec:protocol}

This section gives an overview of the consumer RSP protocol. We explain the protocol participants, trust relations among them, and message flows between the participants. The information has been collected from the GSMA architecture and technical specifications~\cite{RSP-21v2,RSP-22v2} and the GlobalPlatform secure channel protocol specification~\cite{SCP11}, totaling 446 pages. The M2M RSP protocol specification~\cite{RSP-01,RSP-02} was also used as background information.


\subsection{Consumer RSP participants and phases}
\label{sec:ch_protocol_participants}

\begin{figure*}[t]
  \centering
    \includegraphics[width=0.6\linewidth]{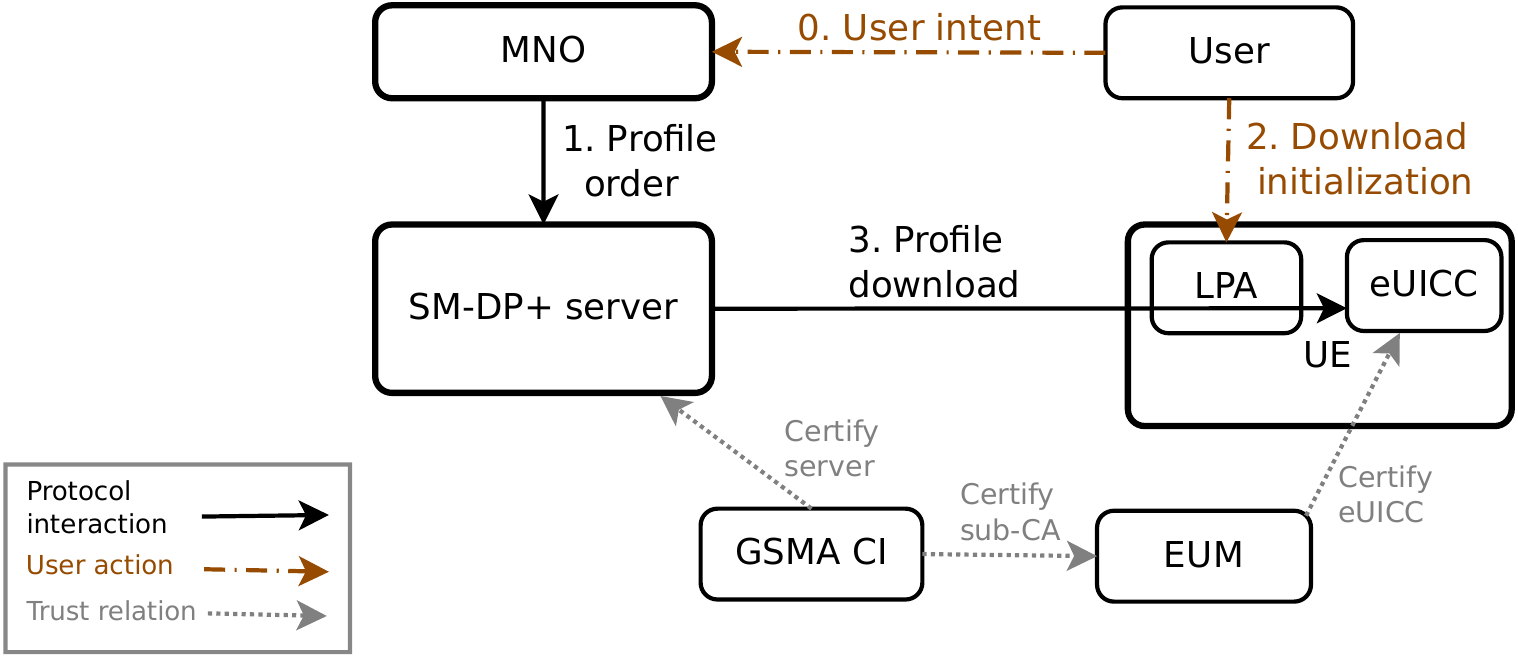}
    \caption{Participants and their interaction in the consumer remote SIM provisioning protocol}
    \Description{The figure shows the main participants, their trust relations, and data flows between the participants.}
    \label{fig:fig-remote-sim-provisioning}
\end{figure*}

Figure~\ref{fig:fig-remote-sim-provisioning} shows the participants of the RSP protocol, their trust relations, user interactions, and the message flow for the profile download. The active participants are:

\begin{enumerate}
  \item \textit{Mobile network operator} (MNO) provides mobile network access to users.
  \item \textit{User equipment} (UE) connects to the mobile network.
  \item \textit{Local profile assistant} (LPA) is a software application that runs on the UE.
  \item \textit{User} owns the UE and has a subscription with the MNO.
  \item \textit{eUICC} is an embedded secure hardware module inside the UE. It stores the \textit{SIM profiles}.
  \item \textit{Subscription manager data preparation} (\SMDP) \textit{server} (or just \textit{server}) is a new entity in the RSP protocol.
\end{enumerate}

To connect to the MNO's mobile network, the UE must have in its eUICC a SIM profile with the necessary credentials. To obtain the SIM profile, the user first contacts the MNO; this can take place on the web, with the help of the LPA application, or offline. The MNO initiates the RSP protocol by sending the profile order to the \SMDP server. The server prepares the SIM profile and makes it available for download. The user then asks the LPA application to download the SIM profile from the server to the eUICC.

We describe the protocol in two parts: (1) the profile ordering and download initialization phases, and (2) the common handshake and profile download phases. Variants of the former are shown in Figures~\ref{fig:fig-default-smdp} and \ref{fig:fig-AC-download}, and the latter is in Figure~\ref{fig:commonhandshake}. Tables~\ref{tab:certificate_notation}--\ref{tab:protocol_events} summarize the notation used in the protocol description.

\subsection{Trust relations and adversary model in the RSP specification}
\label{sec:RSPTrustModel}

In RSP, both the \SMDP server and eUICC trust a common GSMA \textit{certificate issuer} (CI). The CI issues certificates to the \SMDP servers and to \textit{eUICC manufacturers} (EUM). The EUM acts as a subordinate certificate authority and issues certificates to eUICCs. Consequently, the \SMDP server and the eUICC can establish mutual trust using certificate chains that start from a common root authority.

The RSP specification assumes that the CI, \SMDP server, EUM, eUICC, and MNO are all trusted participants. In other words, the adversary has no access to their secret keys and cannot control their operations. In particular, the specification assumes that the adversary cannot obtain a certificate for an arbitrary eUICC identifier and key. The LPA running in the UE is a trusted application that is assumed not to leak any information to the adversary.

The RSP specification mainly considers a \textit{network adversary}, that is, an attacker who can intercept and alter network communications between the RSP participants. The adversary can also trigger a profile download for a small number of eUICCs in mobile devices which the adversary owns.

\begin{table}[p]

\begin{minipage}{0.95\textwidth}
  \centering
  \caption{Certificates in RSP}
  \label{tab:certificate_notation}
  {\small
    \begin{tabularx}{\textwidth}{p{2.6cm}>{\raggedright\arraybackslash}p{1.3cm}>{\raggedright\arraybackslash}p{4cm}>{\raggedright\arraybackslash}X}
    \toprule
    Id, certificate, key & Subject & Purpose & GSMA reference \\
     \midrule
    \CI, \cert{CI}, \SK{CI} & GSMA CI & Root CA & CERT.CI.ECDSA, SK.CI.ECDSA \\
    $S$, \cert{St}, \SK{St} & \SMDP & TLS authentication & CERT.DP.TLS, SK.DP.TLS \\
    \Sa, \cert{Sa}, \SK{Sa} & \SMDP & Server authentication & CERT.DPauth.ECDSA, SK.DPauth.ECDSA \\
    \Sp, \cert{Sp}, \SK{Sp} & \SMDP & Profile binding & CERT.DPpb.ECDSA, SK.DPpb.ECDSA \\
    $U$, \cert{U}, \SK{U} & eUICC & eUICC authentication & CERT.U, SK.U \\
    $\mathit{EUM}$, \cert{\mathit{EUM}}, \SK{\mathit{EUM}} & EUM & eUICC Manufacturer sub-CA & CERT.EUM, SK.EUM \\
    \bottomrule
  \end{tabularx}
  }
\end{minipage}

\bigskip
\begin{minipage}{0.95\textwidth}
  \centering
  \caption{Notations and symbols}
  \label{tab:protocol_notation}
  {\small
  \begin{tabularx}{\textwidth}{p{2.2cm}>{\raggedright}p{6.2cm}X}
    \toprule
    Symbol & Description & GSMA reference \\
    \midrule
    $U$ & eUICC identifier & EID \\
    $S$ & Server domain name & \\
    $\mnoid$ & MNO identifier & \\
    $\UserId$ & User identity authenticated by MNO & \\
    $P$ & SIM profile including the secret key $K_i$ for AKA & \\
    $iccid$ & Unique identifier of the SIM profile & ICCID \\
    $\Iac$ & Matching identifier in activation code & \\
    \SKI{CI} & SubjectKeyIdentifier of \cert{CI} & \\
    $\Seq$ & Sequence counter for profile install notifications &  \\
    $N_U := R$, $N_S := R$ & Random numbers generated by eUICC and server & \\
    $I_t := R$ & Random session identifier generated by server & \\
    $d_U,\; Q_U = d_U \cdot G$ & eUICC ECDHE key components & \\
    $d_S,\; Q_S = d_S \cdot G$ & Server ECDHE key components & \\
    $Z_{US}$ & ECDHE shared secret between eUICC and server & \\
    $\mathrm{KDF}$ & Key derivation function & \\
    $k$, $k'$ & Encryption and integrity keys for server and eUICC & S-ENC, S-MAC \\
    $\enc{k}{M}$ & Encryption of message $M$ with symmetric key $k$ & \\
    $\mac{k'}{M}$ & MAC of message $M$ with symmetric key $k'$ & \\
    $\sign{SK}{M}$ & Signature of message $M$ with private key $SK$ & \\
    $\oid \textrm{ from } \cert{x}$ & Server OID in \cert{Sa} and \cert{Sp} & \\
    $\textrm{subject of } \cert{}$ & Subject identifier to which \cert{} was issued & \\
    \cert{a} $\triangleleft$ \cert{b} & Check that \cert{a} is signed by the subject key of \cert{b} & \\
    $:=$, $=$ & Assignment operator, equality comparison & \\
    $[]$ & \red{Optional parameter, can be null} & \\
    \bottomrule
  \end{tabularx}
  }
\end{minipage}

\bigskip
\begin{minipage}{0.95\textwidth}
  \centering
  \caption{ProVerif events}
  \label{tab:protocol_events}
  {\small
  \begin{tabularx}{\textwidth}{p{4.2cm}X}
    \toprule
    Event and their parameters & Description \\
    \midrule
    AUTHORIZE($\Sp$) & GSMA authorizes the server to issue profiles \\
    OWNER($\UserId,U$) & User is the owner of an eUICC \\
    INTENT($\UserId,\mnoid,U,\Iac$) & User requests a profile from an MNO \\
    ORDER($\UserId,\mnoid,S,U,P,\Iac$) & Server receives an order for a profile from MNO \\
    U0($U,S$) & eUICC begins authentication of the server \\
    U1($U,\Sa,I_t,S$) & eUICC completes authentication of the server \\
    U2($U,\Sa,\Sp,I_t$) & eUICC accepts the server as a profile download entity \\
    U3($U,\Sa,\Sp,I_t,k,P,\mnoid,\Iac$) & eUICC accepts session keys and accepts a profile from the server \\
    S0($\Sa,I_t,S,\mnoid,\Iac$) & Server begins authentication of the eUICC \\
    S1($U,\Sa,\Sp,I_t,\mnoid,\Iac$) & Server completes authentication of the eUICC \\
    S2($U,\Sa,\Sp,I_t,k,P,\mnoid,\Iac$) & Server starts key agreement and profile delivery \\
    S3($U,\Sa,\Sp,I_t,P,S,\mnoid$) & Server accepts profile download notification \\
    \bottomrule
  \end{tabularx}
  }
\end{minipage}

\end{table}

\subsection{Profile ordering and download initialization phases}
\label{sec:ordering_and_initialization}

In the profile ordering phase, an MNO requests a server to prepare a SIM profile. The server then prepares the profile and returns download initialization pointers (e.g.,\ an activation code) to the MNO. In the download initialization phase, the MNO delivers the download initialization pointers to a user and their LPA software to trigger the profile download. The RSP specification supports three different approaches to profile ordering and download initialization: (a) default server, (b) activation code, and (c) subscriber manager discovery server (SM-DS) assisted. For brevity, we model the first two approaches and defer discussion of the third one to Section~\ref{sec:security_SM_DS}.

\begin{figure*}[t]
 \centering
 \includegraphics[scale=0.9]{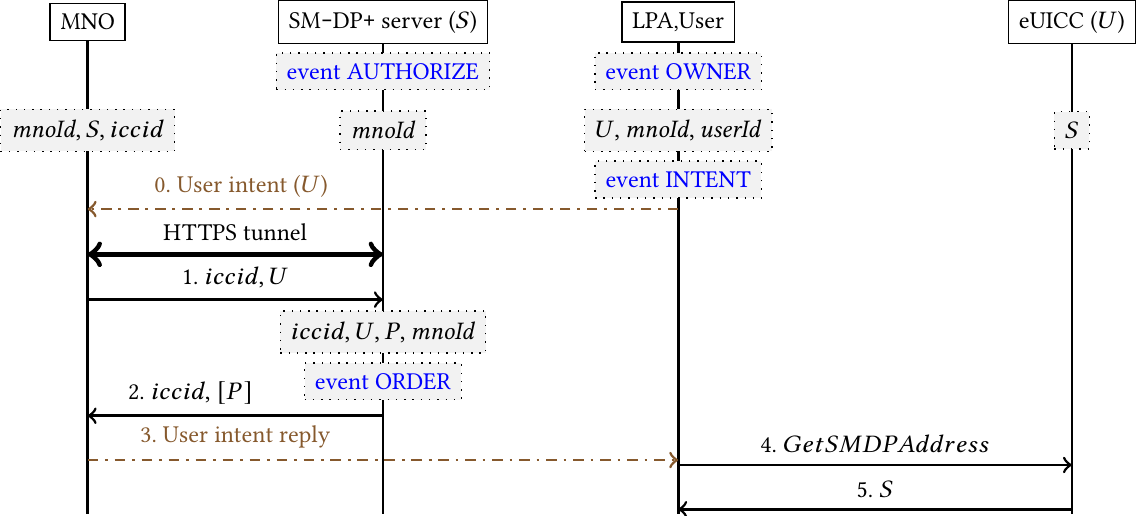}
 \caption{Default-server approach to profile ordering and download initialization: the server stores the eUICC identifier $U$, and the LPA reads the default server domain name $S$ from the eUICC}
 \Description{Message sequence chart that shows the messages between the MNO, server, LPA, and eUICC in the default-server approach.}
 \label{fig:fig-default-smdp}
\end{figure*}

\paragraph{Default server approach} Figure~\ref{fig:fig-default-smdp} shows the default-server approach. The UE manufacturer or reseller, which can be the MNO, pre-provisions the default \SMDP server domain name~$S$ to the eUICC.
\red{When the user intends to get a mobile subscription, the user selects and contacts the MNO either online with the help of the LPA software or with some other method such as going personally to a point of sales. The MNO registers the user identity and the user's eUICC identifier $U$.} The MNO orders a profile for the specific eUICC identifier~$U$ to the default server. Consequently, the server prepares a new SIM profile~$P$ for the eUICC.
The MNO notifies the user, who starts the LPA software in the UE. The LPA retrieves the default \SMDP address~$S$ from the eUICC. The LPA is now ready to move to the common handshake and profile download phases. The profile will be available for download at the address~$S$ only for the eUICC with the certified identity~$U$. 

\begin{figure*}[t]
 \centering
  \includegraphics[scale=0.9]{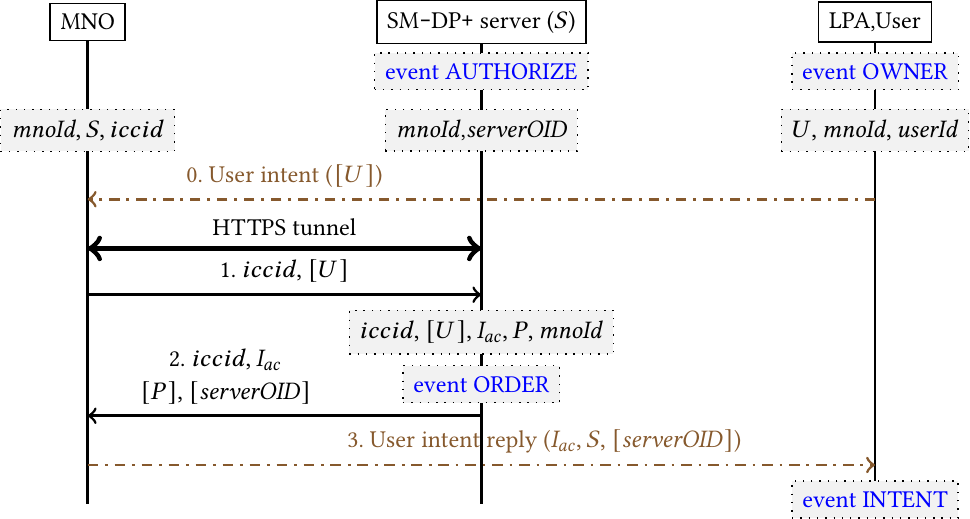}
  \caption{Activation-code approach to profile ordering and download initialization: LPA stores the matching identifier \Iac, the server address $S$, and optionally the server OID; the server optionally stores the eUICC identifier $U$}
  \Description{Message sequence chart that shows the messages between the MNO, server, LPA, and eUICC in the activation-code approach.}
  \label{fig:fig-AC-download}
\end{figure*}

\paragraph{Activation code approach} Figure~\ref{fig:fig-AC-download} shows the activation-code approach. The MNO orders SIM profiles to a server. The server prepares the profiles and makes each profile available with an activation code, which it returns to the MNO. The activation code includes the server domain name~$S$, a matching identifier~\Iac that is unique to the profile, and optionally, a server OID, which is a unique identifier for the \SMDP server. In this approach, the profile preparation can be done either before or after the user contacts the MNO for a new subscription.
When the user intends to get a mobile subscription, the MNO delivers an activation code including $\Iac$ to the user. The user inputs the code into the LPA software, for example, by scanning a two-dimensional bar code. The profile can then be downloaded by anyone who knows the activation code. The matching identifier $\Iac$ in the activation code has to be secret and sufficiently random to be unguessable.

In a variant of the activation-code approach, the user or LPA additionally communicates the eUICC identifier~$U$ to the MNO, which passes it on to the server. In this case, the profile can be downloaded only to the specific eUICC, which must have both the certified identity~$U$ and the matching identifier~\Iac. Pre-establishing the eUICC identifier is practical when the user requests the profile from the MNO with the LPA software and the MNO and \SMDP server are working closely together. It is not convenient to ask the user to communicate the eUICC identifier to the MNO.

In the security analysis, we first assume that the activation code does not include the server OID and that the server does not know $U$. We then explain how these optional features would improve the protocol security.

\paragraph{TLS and out-of-band channels} In all these approaches, the communication between the MNO and the \SMDP server is protected with mutually authenticated TLS~\cite{rescorla2018transport}. The MNO is the TLS client, and the \SMDP server is the TLS server. Both have certificates~\cite{cooper2008rfc}. Trust for this channel is established earlier so that both sides know the other's certified name. 

The user's communication with the MNO may be protected in several different ways: (1) the user goes in person to a shop that represents the MNO and requests a profile; (2) the user logs into an MNO portal online and requests a profile; or (3) the user requests a profile with the help of the LPA or another software application. In the first case, the communication occurs out-of-band, and in the other two cases, over a TLS connection where the MNO is authenticated with a web server certificate. The user must be authenticated as well. The method and strength of the user authentication depend on local regulations and business decisions. In the activation-code approach, the activation code ($\Iac,S,[\oid])$ is communicated to the user and LPA as part of the same protected interaction.

\subsection{Common handshake and profile download phases}
\label{sec:handshake}

\begin{figure*}
 \centering
 \includegraphics[scale=0.9]{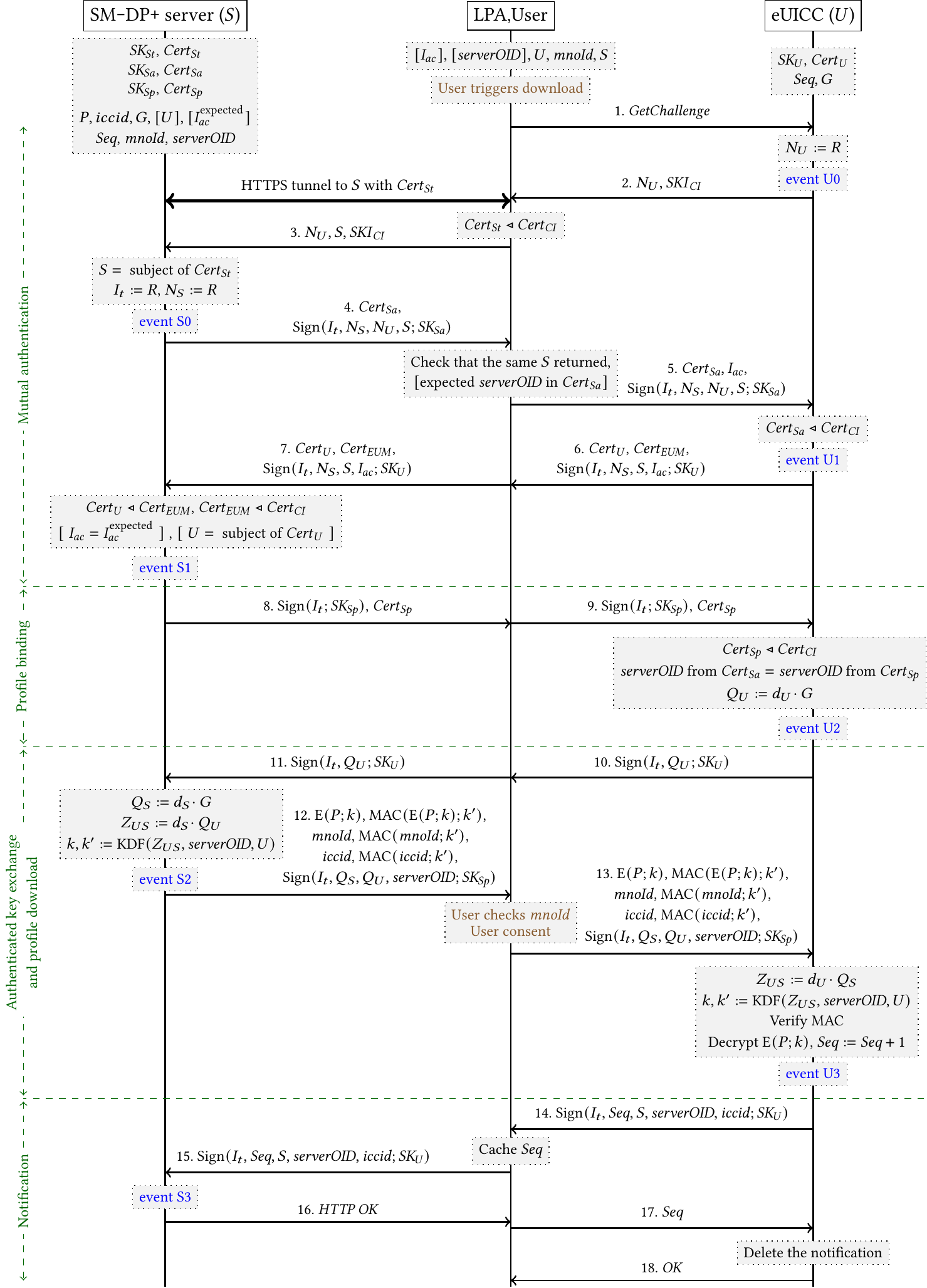}
 \caption{Common handshake and profile download in the RSP protocol}
 \Description{Message sequence chart of the main part of the RSP protocol. The communication endpoints are the server and eUICC. The communication passes through the LPA software on the mobile device.}
\label{fig:commonhandshake}
\end{figure*}

When the profile has been ordered and the download initialized, it is time for the common handshake and profile download phases~\cite[Section~3.1.2]{RSP-22v2}. These are shown in Figure~\ref{fig:commonhandshake}.

There are three main entities involved in these phases: the eUICC, the LPA and user, and the \SMDP server. The certificates used in the handshake are summarized in Table~\ref{tab:certificate_notation}. The server has three different certificates: \cert{St} is the TLS server certificate for the domain name in the server address~$S$. \cert{Sa} is used to authenticate the server to the eUICC, and~\cert{Sp} authorizes the server to issue SIM profiles. The last two certificates have the same subject name, and they contain the same server OID; however, they contain different certificate policy identifiers. \cert{St} may also contain a server OID, but it is not compared with the other two OID values because the certificate is verified by the TLS layer in the client and not by the RSP protocol implementation. The GSMA CI directly issues all the server certificates. The eUICC has a certificate chain consisting of two certificates~\cert{U} and~\cert{EUM}. The latter is an equipment manufacturer certificate issued by the GSMA CI, and the former is the eUICC certificate issued by the manufacturer to the eUICC identifier~$U$.

At the beginning of the profile download phase, the server always has the profile~$P$ and the associated MNO identifier \mnoid. The server also has the receiving eUICC identifier~$U$ or a matching identifier of an activation code~\Iac, or both. The LPA software knows the server address~$S$, and it may have a matching identifier \Iac and the server OID.

We will now explain the common handshake and profile download phases in Figure~\ref{fig:commonhandshake} step by step.

\paragraph{TLS tunnel} The protocol takes place inside a TLS tunnel from the LPA to the \SMDP server, which is authenticated with the server certificate~\cert{St}. The TLS client is not authenticated. The ways in which the LPA can learn the server domain name~$S$ were discussed above in Section~\ref{sec:ordering_and_initialization}.

\paragraph{Mutual authentication (messages 1--7)} The eUICC and the server authenticate each other inside the TLS tunnel. In the base case, these endpoints do not have prior knowledge of each other's identities $U$ and $Sa$. The eUICC indicates in message 2 its preferred CI root key with~\SKI{CI}. This key identifier also determines the elliptic curve that will be used for all the certificates, signatures, and ECDH key exchange throughout the protocol. The server accepts any valid eUICC certificate chain~\cert{U},~\cert{EUM} that originates from the root key, and the eUICC accepts any valid server certificate~\cert{Sa} issued by the same root key. The nonces~$N_S$ and~$N_U$ ensure the freshness of the mutual authentication. During the authentication, the server selects a fresh session identifier~$I_t$, which will be used to identify the ongoing session throughout the rest of the protocol.

The server includes the server domain name~$S$ from its TLS certificate in the signed message 4. Before passing the message to the eUICC, the LPA verifies that~$S$ matches the server name to which it created the TLS tunnel. This provides a level of channel binding between the TLS tunnel and the inner authentication. If the LPA knows the server OID beforehand, it verifies that the server certificate~\cert{Sa} contains the correct~\oid. \red{On the other hand, the specification does not require the server to compare $S$ in message 7 with its own name. This could be an accidental omission rather than intended behavior. We nevertheless model the protocol as specified.} 

In the activation-code approach, the LPA gives the matching identifier~\Iac to the eUICC in message 5, and the eUICC includes it in the signed message 6--7 to the server. The server then selects the correct profile for the eUICC based on the matching identifier. In the default-server approach, the LPA sets the matching identifier to a null value and the server selects the profile based on the eUICC identifier $U$ in~\cert{U}.

\paragraph{Profile binding (messages 8--9)} After the mutual authentication, the server checks that it has a profile available for download to the eUICC. If it does, it signs the session identifier~$I_t$ with the profile binding key and sends the signature together with the profile-binding certificate~\cert{Sp} to the eUICC. The eUICC verifies the signature and certificate, and it also checks that the two server certificates~\cert{Sa} and~\cert{Sp} contain the same server OID. This step assures the eUICC that the previously authenticated server is an authorized \SMDP server.

\paragraph{Authenticated key exchange (messages 10-13)} The key exchange is based on elliptic-curve key agreement (ECKA)~\cite{SCP11,BSITR0311}. The eUICC and server generate ephemeral elliptic-curve Diffie-Hellman (ECDHE) key components $d_U,\; Q_U = d_U \cdot G$ and $d_S,\; Q_S = d_S \cdot G$, respectively. The eUICC and server exchange the public components in messages 10--13 and then compute the shared secret $Z_\mathit{US} = d_S \cdot Q_U = d_U \cdot Q_S$. From this shared secret, they generate session keys $k$ and $k'$ with the key derivation function (KDF)~\cite[Section 4.3.3]{BSITR0311}. Other inputs to KDF are the server OID and the eUICC identifier~$U$. The key exchange is authenticated by the signatures and certificates in messages 10-11 and 12-13. In the following download phase, $k$ will be used for data encryption and $k'$ for data authentication.

\paragraph{Profile download (messages 12-13)} This phase overlaps with the last message of the authenticated key exchange. The server performs authenticated encryption~\cite[Section 4.1.3]{RSP-02} of the profile~$P$. That is, it first encrypts the profile with the key $k$ and then appends a message authentication code (MAC) computed with the key~$k'$. The server sends the encrypted profile $\enc{k}{P}$ and the MAC to the LPA. It also sends the \mnoid (and other metadata that is unimportant for the current discussion) protected with a separate MAC.

The \mnoid identifies the mobile network operator that asked the server to prepare the profile. The LPA displays \mnoid to the user, who should check that it refers to the intended operator. 
Once the user has approved the operator, the LPA forwards the information to the eUICC. Consequently, the eUICC computes the session keys~$k,~k'$, checks the MACs, and decrypts the received profile $P$. The eUICC can now take the profile $P$ into use.

\paragraph{Notification (messages 14-17)} After accepting the downloaded profile, the eUICC deletes the session keys~$k,~k'$ and notifies the LPA with a signed notification message. The LPA may deliver the notification to the server immediately or after a delay. The signed message contains the server address and OID as well as a sequence number~\Seq. For this, the eUICC maintains a sequence counter, which it stores in non-volatile memory and increments by one for each new notification. The server checks that the received sequence number is greater than the previous one received from the same eUICC. The server then responds with \textit{HTTP OK} to the LPA. The LPA asks the eUICC to delete the notification from its queue, and the eUICC responds with \textit{OK} status to the LPA.

\section{Formal model of RSP security}
\label{sec:formalmodel}

\begin{figure*}[t]
  \centering
  \includegraphics[width=0.5\linewidth]{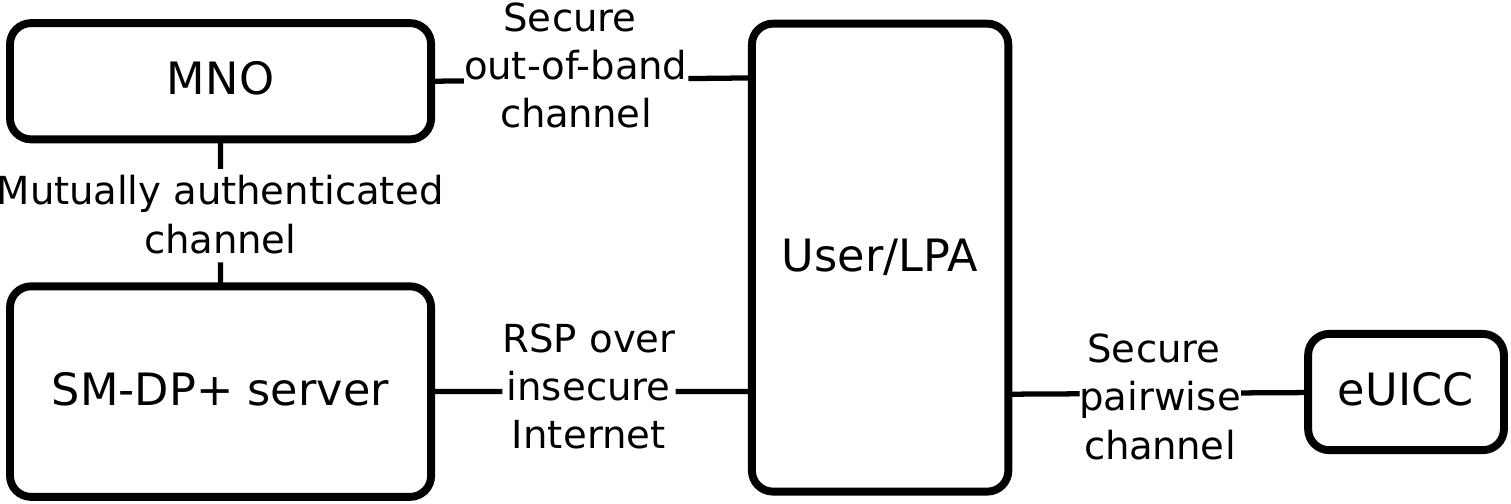}
  \caption{Participants and channels in the protocol model}
  \Description{In the simplified view, the four main participants are the MNO, the server, and the client, which comprises the user and the LPA software together, and the eUICC.}
  \label{fig:fig-entity-interface}
\end{figure*}

This section gives an overview of the ProVerif model\footnote{The model is available at \url{https://github.com/peltona/rsp_model}},
defines the partial compromise scenarios, and discusses some of the abstractions and security assumptions made in the model. Figure~\ref{fig:fig-entity-interface} shows the participants and communication channels in the model.

\subsection{Model overview}
\label{sec:formalmodel_model}

\paragraph{Participants} We explicitly model four participants of the RSP protocol as ProVerif processes: (1) MNO, (2) \SMDP server (3) user and UE with LPA software, and (4) eUICC. The roles of the user, UE, and LPA have been merged into a single process because their integrity is mutually dependent: on one hand, the user could choose a hacked phone as the UE or install modified LPA software, and on the other hand, the faithful execution of the user's intent depends on both the LPA software and UE hardware. In comparison, even though the eUICC is in the user's possession, it is a hardware module with a security perimeter that protects its local integrity and secrets.

\paragraph{Communication channels} We model the communication channels between the participants as follows. \red{The \textit{MNO-to-server} channels is modeled as private channel to which the adversary has no access. It is implemented as a mutually-authenticated TLS channel that does not use the web PKI. Therefore, we can assume that only authenticated MNOs have access to this channel. The \textit{User/LPA-to-MNO} communication takes place out-of-band and is also modeled as a private channel.} Some of the partial compromise scenarios represent failures of this channel, such as identity fraud where the adversary is able to impersonate the user on the User/LPA-to-MNO channel.
The \textit{User/LPA-to-server} channel is the main target of our investigation. We model it as a public channel that is vulnerable to the network adversary. The server-authenticated TLS tunnel on the User/LPA-to-server channel is modeled explicitly. Finally, the \textit{User/LPA-to-eUICC} channel is a pairwise secure channel inside the UE.

\paragraph{Events} Table~\ref{tab:protocol_events} lists the events emitted by the protocol participants that will be used for specifying the security goals. The same events occur in Figures~\ref{fig:fig-default-smdp}--\ref{fig:commonhandshake}.
The event AUTHORIZE indicates the legitimacy of the \SMDP server, and OWNER indicates the user's ownership of the device that contains the eUICC. INTENT models the user's decision to obtain a SIM profile from the MNO, which triggers the RSP execution. The events OWNER and INTENT are linked together by a parameter \UserId, which is an abstract representation of the authenticated user identity and does not appear in the RSP messages. The server emits the event ORDER when it receives the profile order from the MNO. The remaining events U0--U3 and S0--S3 signal the lockstep progress of the \SMDP server and the eUICC through the common handshake and profile download phases. S2 and U3 respectively indicate the completion of the critical parts of the protocol at the server and eUICC.

\subsection{Assets that require protection}
\label{sec:ch_model_asset}

The focus of the model is on the confidentiality and integrity of the SIM profile. The profile is initially known only to the \SMDP server and MNO. After the eUICC downloads the profile, it is assumed that the server forgets the data and the eUICC stores it securely. With these assumptions, the primary goal of the RSP protocol is to establish the profile as an authenticated shared secret between the MNO and the intended eUICC.

The profile is modeled as an opaque block of data $P$. That is, the contents of the profile are not modeled explicitly. For simplicity, the reader may want to think of the profile as the shared key $K_i$ that will be used for the AKA procedure~\cite{3gpp_3g_2020,arkko2006extensible} between the UE and the MNO to authorize mobile network access. The profile also contains other data, such as the subscriber and MNO identifiers, which require integrity protection.

\subsection{Adversary model and partial compromise scenarios}
\label{sec:compromise_scenarios}

The Dolev-Yao adversary model, which is built into ProVerif, assumes that the endpoints are trusted while the underlying network is untrusted. That is, the communication network is assumed to be under the adversary's control.

Given the use of TLS tunnels for network communication and out-of-band authentication between the user and the MNO, we do not expect major security failures against the base-case network adversary. Nevertheless, it is necessary to verify that the TLS tunnels and the protocols executed inside these tunnels combine to provide the desired security properties.

More importantly, we extend the network adversary in the following scenarios by modeling the compromise of selected endpoints and channels:\\

\newcommand{\scenarioitem}[1]{\refstepcounter{enumi}\item[\textbf{\theenumi. #1:}]}
\fbox{\parbox{0.94\textwidth}{%
\begin{enumerate}[align=left,leftmargin=1.4cm,itemsep=0.5\parsep,label={\textbf{\theenumi}}]
  \scenarioitem{base-case} All participants are honest. The network between the LPA and \SMDP server is untrusted.
  \scenarioitem{server} The adversary has all the private keys of the \SMDP server.
  \scenarioitem{eUICC} The adversary has the private key \SK{U} of the honest user's eUICC.
  \scenarioitem{LPA} The adversary is the user who controls the LPA, or compromised LPA software misleads the user.
  \scenarioitem{2nd server} The intended \SMDP server is honest, but there exists another compromised server.
  \scenarioitem{2nd eUICC} The adversary has compromised the private key of the eUICC in its own device.
  \scenarioitem{2nd MNO} The intended MNO is honest, but there exists another compromised MNO.
  \scenarioitem{order as user} The adversary impersonates the honest user when ordering a profile.
  \scenarioitem{order for eUICC} The adversary requests a profile for the honest user's eUICC.
  \scenarioitem{AC leak} The adversary learns the activation code associated with an ordered profile.
  \scenarioitem{AC spoof} The adversary tricks the LPA or user to use the wrong activation code.
\end{enumerate}}}\\

\red{The partial compromised scenarios are based on the model structure in Figure~\ref{fig:fig-entity-interface}. We consider the potential compromise of each entity. We then differentiate between the failure of that entity as an intended participant of the protocol (scenarios 2-4) and as an outsider (scenarios 5-7). However, a compromise scenario for the user's chosen MNO is not included because no security for cellular communication can be expected with a compromised operator. An attacker-controlled 2nd LPA is part of the base model and thus not a separate scenario. We additionally consider the compromise of the User/LPA-to-MNO channel integrity (scenarios 8-9, 11) and secrecy (scenario 10) to cover the potential mistakes which the user could make as an intermediary on the channel.}

In scenarios 2--3, one of the main communication endpoints, either the \SMDP server or eUICC, is compromised. This is modeled by leaking the endpoint's private keys to the adversary. In practice, the server may have been hacked, or the eUICC keys may have been compromised in the supply chain. A serious security failure is expected in these scenarios. They are included in the analysis because we can learn about the failure modes --- and also as sanity checks for the model.

In scenario 4, the adversary is the mobile device user. They can install modified LPA software and even UE firmware, but unlike in the previous scenario, they have not compromised the integrity of the eUICC hardware inside the UE. This scenario also covers the case where an honest user performs all their actions, including the profile ordering, with compromised LPA software.

Scenarios 5--7 represent situations where the adversary has compromised one \SMDP server, eUICC, or MNO and uses that to attack RSP between honest participants. The existence of the compromised MNO should not influence the security of other MNOs and their customers --- not even if they use the same \SMDP server. Likewise, the existence of a compromised server should not influence the security of those who use a different server. Also, an advanced adversary could compromise the integrity of a small number of eUICCs and extract their private keys in the laboratory, but that should not affect the security of other eUICCs. \red{The compromised entities are modeled primarily by leaking their private keys. However, some communication in the ProVerif model is represented with private channels. In these cases, we explicitly model the compromised entity as a proxy between the attacker and the private channel.}

Scenarios 8--11 model four different types of fraud during the profile ordering. In scenario 8, the adversary can impersonate the honest user in the ordering phase, and in scenario 9, the adversary knows the identifier of the honest user's eUICC, which enables it to order profiles for the eUICC. In scenario 10, the adversary spies the activation code from a careless user, and in scenario 11, the adversary tampers with the activation code that is given to the user or LPA. While the adversary in these scenarios is relatively weak, these scenarios are likely to occur in the real world.

In all the scenarios including the base case (scenario 1), the adversary has its own user identity and eUICC, for which it can order profiles with the RSP protocol. This represents the adversary's ability to purchase a mobile phone and a mobile network subscription for it.

We model all these scenarios \textit{with and without TLS tunnel} between the LPA and \SMDP server. The primary reason is to investigate how important the TLS layer is for the security of RSP. Another reason is that the TLS server is exposed directly to the Internet and its private TLS key \SK{St} is thus susceptible to hacking. This is more likely than the full compromise of the server (scenario 2) because the application keys \SK{Sa} and \SK{Sp} are typically stored in a secure hardware module or in a backend server.

When an entity's private key leaks in scenarios 5-7, the model emits a marker event, e.g.~\event{CompromiseCert}{U} in the case of a compromised eUICC. The formal verification goals will use these events to differentiate between the compromised entity and honest ones of the same type. Scenarios 8-9 emit similar marker events to differentiate between dishonest and honest orders.

We generate separate models for each of the two profile ordering and download initialization approaches in the relevant partial compromise scenarios. In order to understand the significance of the TLS tunnel between the UE and server, we generate separate model variants with TLS and without. The result is 40 different ProVerif models.

\subsection{Abstractions made in the model}
\label{sec:model_abstractions}

In the symbolic protocol model, cryptographic primitives including signature, key agreement, and encryption algorithms are modeled as simple rewriting rules without considering their cryptographic implementation.
Some of the message fields are abstracted away: we omit eUICC capabilities, cryptographic algorithm suites, message lengths, profile contents, and error codes. The activation code never expires in our model. Also, the eUICC certificate in the model is issued directly by the CI (and not by the EUM), so that we avoid modeling the two-certificate chain.

ProVerif supports unlimited process replication for verifying protocols with an unlimited number of participants and an unlimited number of sessions between them. All the positive verification results reported in this paper were verified with \textit{unlimited participants} and \textit{unlimited sessions}. To complete some verification tasks, we had to simplify the model by combining the User/LPA and eUICC into one process. This does not limit the generality of the results because the optimization was necessary only in cases where ProVerif found an attack but needed help in reproducing the attack trace.
\section{Formalization of the security goals}
\label{sec:formalgoals}

This section formalizes the security goals of the RSP protocol. The RSP specification~\cite[sec.~2.6.2]{RSP-22v2} defines some explicit security requirements, and implicit requirements exist throughout the specification. We derive precise formal security goals from both types of requirements. We then use ProVerif to verify the goals against the formal protocol model described above. In the following, the message numbers and events refer to Figures~\ref{fig:fig-default-smdp}--\ref{fig:commonhandshake}.

There are both authentication and secrecy goals. The \textit{authentication goals} are expressed as correspondences between local events in the participating processes. The event parameters represent local variable values in the process where the event occurs. The \textit{secrecy goals} in sections~\ref{sec:AuthD} and \ref{sec:overall_goals} state that a session key or profile accepted by one honest participant must not be known to the adversary. The events and their full parameters are listed in Table~\ref{tab:protocol_events}. In the security goals below, we use the underscore $\freevar{x}$ to denote event parameters that are unused or unimportant for the security goal in question. In the formal queries, each $\freevar{x}$ on the left side of the $\IMPLY$ can be read as a universally quantified wildcard variable and, on the right side, as an existentially quantified one.

The security goals can also be divided into server-side and client-side ones. The
\textit{server-side} goals formalize the assurance that the server gets from the protocol execution, including authentication of the client; they can be recognized by having a server event to the left of $\IMPLY$. The \textit{client-side} goals express the assurance that the client gets, including authentication of the server; they have a client event to the left of$\IMPLY$.

\subsection{Mutual authentication} 
\label{sec:AuthA}

The first two authentication goals are based on the following explicit authentication requirements from the RSP specification~\cite{RSP-22v2}:

\quotebox{The Server (the entity providing the function, e.g.\ \SMDP) SHALL be authenticated first by the Client. Authentication SHALL include the verification of a valid Server Certificate signed by a GSMA CI.}{}
\quotebox{The Client SHALL be authenticated by the Server in a second step. In case the Client is the eUICC, authentication SHALL include the verification of a valid eUICC and EUM Certificate signed by a GSMA CI.}{}
\quotebox{\red{The eUICC, as a Client, SHALL not generate any signed material before having authenticated the Server.}}{}
\quotebox{\red{On the basis of authentication, the Server SHALL always check that the requesting Client is authorized before delivering the requested function execution.}}{}

The mutual entity authentication takes place in messages 1--7 (see Figure~\ref{fig:commonhandshake}). \red{Client authorization is based on its certificate chain and on the previous order of a profile with the matching eUICC identity $U$ or activation code $\Iac$}. These properties are formalized all together as the following correspondence properties:
\begin{equation}\tag{Auth A}
  \event{U1}{U,Sa,I_t,S} \IMPLY
  \injevent{U0}{U,[S]}
  \AND \injevent{S0}{Sa,I_t,S,\freevar{\mnoid},\freevar{\Iac}}
\end{equation}
\begin{equation}\tag{Auth B}
  \begin{aligned}
  \event{S1}{U,\Sa,\freevar{\Sp},I_t,\mnoid,\Iac} \IMPLY
  &\injevent{S0}{\Sa,I_t,S,\mnoid,\Iac}
  \AND \injevent{U1}{U,\Sa,I_t,S}
  \end{aligned}
\end{equation}

Auth A and Auth B above are \textit{stepwise} goals in the sense that they require correspondence over single steps of the protocol execution. Transitively, these properties say that, if the client or the server accepts the mutual authentication (event U1 or S1, respectively), then both sides must have started the authentication (both events U0 and S0).

In addition to the stepwise goals, it is possible to specify transitive correspondence between the latest event and initial events that triggered the protocol execution. We have chosen to include here one such transitive goal, Auth~B$'$, because it provides some interesting verification results.
\begin{equation}\tag{Auth B$'$}
  \begin{aligned}
  \event{S1}{U,\Sa,\freevar{\Sp},I_t,\mnoid,\Iac} &\IMPLY
  \event{OWNER}{\UserId,U} \\[-3pt]
  &\AND \event{INTENT}{\UserId,\mnoid,U,\Iac} \\[-3pt]
  &\AND \injevent{ORDER}{\UserId,\mnoid,\freevar{S},[U],\freevar{P},\Iac}
  \end{aligned}
\end{equation}
Auth B$'$ specifies a correspondence between the end of the authentication step and the initial user intent and profile order. The correspondence with ORDER is injective, meaning that event $S1$ should occur at most once for each profile order received by the \SMDP server.
The user identity \UserId and the concept of ownership do not occur in the RSP specification; they are needed in the formal model to express the user's intent to download a profile for a specific eUICC. In the default-server approach, the ORDER event in the server is associated with an eUICC identifier ($[U]$ in Auth B$'$ denotes U), while in the activation-code approach, the order is not made for a specific eUICC ($[U]$ in Auth B$'$ denotes \textit{null}). In the activation-code approach, the server must accept only the \Iac value registered in the profile ordering phase. In the default-server approach, there is no such value (\Iac = \textit{null}).

The advantage of formal verification is that it forces us to be precise about the security goals. Above, we had to precisely state what knowledge each participant has at the events, which leads us to several important observations.
The formalization makes it explicit that the initial client event U0 does not depend on the server OID or certificate \Sa, and that the initial server event S0 does not depend on the client identity $U$. This draws our focus onto the fact that, in most deployments, neither side knows the identity of the other at the start of the common handshake phase. Instead, the client is willing to accept any server OID that is certified by the GSMA CI, and the server is willing to accept any client identity $U$ that has a certificate chain with GSMA CI as the root. If we included the identifiers as parameters in the initial events in Auth A and Auth B, we would not be able to verify these goals against the protocol model --- or more accurately, we could verify them only against the variant of the model where the identities are communicated through a secure channel in the profile download and initialization phases.

\red{In scenarios 5--7, the compromised entity is not the intended server, eUICC, or MNO, respectively. When analyzing these scenarios, we
exclude results where the compromised entity is an intended participant in the protocol run. Also, we exclude attacks that only violate the security for an attacker-owned eUICC. In the model, this is done by adding disjunctive terms to the right side of the correspondence queries to force ProVerif to ignore these situations that are not real attacks.}

\subsection{Profile binding} 
\label{sec:AuthC}

\red{Server authorization is not explicitly stated as a goal in the RSP specification but it is implied by the concept of profile binding.} After the mutual entity authentication, the server proves to the eUICC that it is authorized to deliver a profile by signing the session identifier~$I_t$ with its GSMA-issued profile binding certificate~\cert{Sp}. We formalize the binding as the correspondence Auth C, which links the subject identity \Sp in the profile binding certificate to the previously authenticated server identity \Sa. The formal goal says that, if the eUICC accepts the two certificates together, \Sp must be an authorized server, and the server must have intended the two certificates to be used together in this session. The injective correspondence indicates freshness of the binding.
\begin{multline}\tag{Auth C}
  \event{U2}{U,\Sa,\Sp,I_t} \IMPLY \\[-3pt]
  \event{AUTHORIZE}{\Sp}
  \AND \injevent{U1}{U,\Sa,I_t,\freevar{S}}
  \AND \injevent{S1}{U,\Sa,\Sp,I_t,\freevar{\mnoid},\freevar{\Iac}}
\end{multline}

In the protocol, the profile binding is implemented so that the eUICC compares the server OID between the profile binding certificate in message 9 and the previously-authenticated server certificate. The freshness depends on rather intricate reasoning. Messages 8--9 do not contain the client nonce $N_U$ or any other fresh value that originates from the client. Instead, the client reasons that the server always chooses a fresh $I_t$ in the mutual authentication step (messages 4--5), which then proves the freshness of messages 8--9 to the client. This kind of indirect reasoning complicates the security analysis and formal verification. Nevertheless, ProVerif is able to verify Auth C without being given any hint about the reasoning.

The process of formalizing the security goals often leads us to question the protocol design details and suggests potential optimizations. While RSP includes the fresh profile binding step (messages 8--9) in each session, an alternative would be to view the ability to provision SIM profiles as a persistent property of the server. By issuing the profile binding certificate \cert{Sp} with the same server as the server certificate \cert{Sa}, GSMA attests that the server is authorized to provision SIM profiles as long as both certificates remain valid. Seen this way, the profile-binding messages do not need to be signed or associated with a specific session. Thus, messages 8--9 could be omitted, and the profile binding certificate could be attached to one of the other messages.

\subsection{Authenticated key exchange and profile download}
\label{sec:AuthD}

\red{The security goals in this section are based on the following requirement from the RSP specification~\cite{RSP-22v2}:}
\quotebox{\red{The communication SHALL be origin authenticated, as well as integrity and confidentiality protected.}}{}

The eUICC and \SMDP server perform an ephemeral ECDH key exchange in messages 10--13. To authenticate the key exchange, they sign the messages with the certificates that were previously exchanged. Note that the server has now switched to using the profile binding certificate~\cert{Sp} and the identity \Sp in it. The following formal security goals state that, if either side accepts the key exchange results, both sides must have started the key exchange. 
\begin{equation}\tag{Auth D}
  \begin{aligned}
    \event{S2}{U,\Sa,\Sp,I_t,\freevar{k},P,\mnoid,\Iac} \IMPLY& \\[-3pt]
    \injevent{S1}{U,\Sa,\Sp,I_t,\mnoid,\Iac}&
    \AND \injevent{U2}{U,\Sa,\Sp,I_t}
  \end{aligned}
\end{equation}
\begin{equation}\tag{Auth E}
  \begin{aligned}
    \event{U3}{U,\Sa,\Sp,I_t,k,\freevar{P},\mnoid,\Iac} \IMPLY& \\[-3pt]
    \injevent{U2}{U,\Sa,\Sp,I_t}&
    \AND \injevent{S2}{U,\Sa,\Sp,I_t,k,\; \freevar{P},\mnoid,\Iac}
  \end{aligned}
\end{equation}

The goal of the key exchange is to create a fresh session key for the profile download. The key exchange neither has nor needs many of the properties that a more generic key exchange protocol would have. As is apparent from Auth D, there is no key confirmation~\cite{fischlin2016key} for the server ($k$ is not a parameter in the events S2 and U2). Auth E, on the other hand, requires key confirmation for the client. In the protocol specification, it is implemented by computing two message authentication codes for messages 12-13 with the related session key $k$ and $k'$.
\red{Because the two keys are derived from the same inputs, we model them as just one value $k$ that is used for both purposes.}

The secrecy of the session key is expressed with the following two queries.
\begin{equation}\tag{Sec W}
  \attacker(k) \AND \event{S2}{U,\Sa,\freevar{\Sp},\freevar{I_t},k,\freevar{P},\mnoid,\freevar{\Iac}}
  \IMPLY \false
\end{equation}
\begin{equation}\tag{Sec X}
  \hspace{-1pt}
  \attacker(k) \AND \event{U3}{U,\Sa,\freevar{\Sp},\freevar{I_t},k,\freevar{P},\mnoid,\freevar{\Iac}} \\
  \IMPLY \false
\end{equation}

Sec W and Sec X state that the adversary must not know a session key that is accepted by the server or eUICC. Critically, this secrecy property must hold even when the eUICC is in an adversary-owned device, i.e., when the LPA and user are untrusted.

\red{Forward secrecy~\cite{diffie1992authentication} in a key agreement assures that session secrets will not be compromised even if the private keys of the participants are compromised later on. The RSP specification requires forward secrecy~\cite{RSP-22v2}:}
\quotebox{\red{Session keys SHALL be generated using Perfect Forward Secrecy.}}{}

\red{We verify forward secrecy for the session keys in the standard way by leaking the long-term secrets of all the honest entities after the completion of the protocol run. Sec W and Sec X should continue to hold. Since RSP uses ECDH with fresh private parameters in each session, we expect forward secrecy to hold whenever Auth D and Auth E are true.}

In messages 12--13, the eUICC finally receives the SIM profile $P$ and the associated metadata \mnoid. The formal security goal below refines Auth E by binding the profile P to the events.
\begin{equation}\tag{Auth F}
  \begin{aligned}
    \event{U3}{U,\Sa,\Sp,I_t,k,P,\mnoid,\Iac} &\IMPLY \\
    \injevent{U2}{U,\Sa,\Sp,I_t}
    &\AND \injevent{S2}{U,\Sa,\Sp,I_t,k,P,\mnoid,\Iac}
  \end{aligned}
\end{equation}

After the events S2 and U3, the protocol has reached its main objective of SIM profile download. At this point, the mobile user can start accessing the mobile network services. Clearly, we should look at the transitive correspondences from S2 and U3 to the initial events U0 and S0 and to the preceding INTENT and ORDER. Before doing that, we will briefly discuss the final messages of the protocol.

\subsection{Profile install notification}
\label{sec:AuthG}

The client acknowledges the successful download and installation of the SIM profile by sending a profile install notification to the server (messages 14-15). The formal security goal below states that the server should only accept the notification if it previously sent the profile to the eUICC and if the eUICC has accepted the profile.
\begin{equation}\tag{Auth G}
  \begin{aligned}
    \event{S3}{U,\Sa,\Sp,I_t,P,\freevar{S},\mnoid} &\IMPLY
    \event{S2}{U,\Sa,\Sp,I_t,\freevar{k},P,\mnoid,\freevar{\Iac}} \\[-3pt]
    &\AND
    \event{U3}{U,\Sa,\Sp,I_t,\freevar{k},P,\mnoid,\freevar{\Iac}} \\[-3pt]
    &\AND \event{OWNER}{\UserId,U} \\[-3pt]
    &\AND \event{INTENT}{\UserId,\mnoid,U,\freevar{\Iac}} \\[-3pt]
    &\AND \event{ORDER}{\UserId,\mnoid,\freevar{S},[U],\freevar{P},\freevar{\Iac}}
  \end{aligned}
\end{equation}

The RSP specification states that ``Seq protects against replay attacks''~\cite{RSP-22v2}. \red{We do not model the sequence number for two reasons. First, the session identifier $I_t$ is sufficient for freshness, as can be verified with an injective version of Auth G. Second, the replay protection is actually unnecessary.} The eUICC includes~\Seq in the notification it sends after successfully installing, enabling, disabling, or deleting a profile. The replay protection is needed when enabling and disabling profiles so that the server knows the current profile state. Installing and deleting, on the other hand, can occur only once for each profile, and replays of the success notification would not cause any confusion. \red{To emphasize this observation, we have presented a non-injective version of Auth G above.}

The notification response (message 16) contains only the \textit{OK} status. The LPA then tells the eUICC to delete the notification from its message queue. These messages are not critical for the security of RSP, and they have no cryptographic end-to-end protection. We omit the formalization but note that the security depends entirely on the TLS tunnel and the integrity of the LPA.

\subsection{Security goals for the full protocol}
\label{sec:overall_goals}

After the stepwise security goals, it is now time to look at the goals for the RSP protocol as a whole. To summarize the outcome of the common handshake and profile download (messages 1--13), we verify an injective correspondence between the eUICC accepting the profile in event U3 and the initial events U0 and S0.
\begin{equation}\tag{Auth I}
  \event{U3}{U,\Sa,\freevar{Sp},I_t,k,P,\mnoid,\freevar{\Iac}} \IMPLY
    \injevent{U0}{U,\freevar{S}}
    \AND \injevent{S0}{\Sa,I_t,\freevar{S},\mnoid,\freevar{\Iac}}
\end{equation}

This formula states that, if the eUICC accepts a profile, it must have started a matching common handshake. It can be verified with any subset of the MNO identifier \mnoid, server domain name $S$ (in the default-server approach), and activation code \Iac (in the activation-code approach) as parameters in the events. We included only \mnoid above because it must match the expected value while the other parameters no longer have significance after the protocol has completed. We will verify Auth I directly with ProVerif. Interestingly, the same result can be obtained as the transitive closure of the stepwise formulas (Auth F, Auth D, and Auth B).
Above, we have focused on the common handshake and profile download phases. The following two goals summarize the overall objectives of RSP also including the profile ordering and download initialization phases. The events S2 and U3 respectively indicate the server sending the profile and the eUICC accepting it. The correspondences are from these events all the way back to the initial user intent and profile order.
\begin{equation}\tag{Auth J}
  \begin{aligned}
  \event{U3}{U,\Sa,\freevar{Sp},I_t,\freevar{k},P,\mnoid,\Iac} &\IMPLY
  \event{OWNER}{\UserId,U} \\[-3pt]
  &\AND \event{INTENT}{\UserId,\mnoid,U,\Iac} \\[-3pt]
  &\AND \injevent{ORDER}{\UserId,\mnoid,\freevar{S},[U],P,\Iac}
  \end{aligned}
\end{equation}
\begin{equation}\tag{Auth K}
  \begin{aligned}
  \event{S2}{U,\Sa,\freevar{Sp},I_t,\freevar{k},P,\mnoid,\Iac} &\IMPLY
  \event{OWNER}{\UserId,U} \\[-3pt]
  &\AND \event{INTENT}{\UserId,\mnoid,U,\Iac} \\[-3pt]
  &\AND \injevent{ORDER}{\UserId,\mnoid,\freevar{S},[U],P,\Iac}
  \end{aligned}
\end{equation}

Auth K builds on the earlier goal Auth B$'$, and we will find that these two goals fail in the same situations.

The ultimate secret that requires protection in RSP is the SIM profile $P$. This is because the profile includes the secret shared key that will be used for authentication between the mobile device and the MNO.
The following formal security goals state that a profile that has been sent by the server or accepted by the eUICC must not be known to the adversary.
\begin{equation}\tag{Sec Y}
  \hspace{-10pt}
  attacker(P) \AND
  \event{S2}{U,\Sa,\freevar{\Sp},\freevar{I_t},k,P,\mnoid,\Iac}
  \IMPLY \false
\end{equation}
\begin{equation}\tag{Sec Z}
  \hspace{-10pt}
  attacker(P) \AND
  \event{U3}{U,\Sa,\freevar{\Sp},\freevar{I_t},k,P,\mnoid,\Iac}
  \IMPLY \false
\end{equation}

\section{Verification results}
\label{sec:results}

This section describes the results from verifying the security goals formalized in Section~\ref{sec:formalgoals} against the ProVerif protocol model described in Section~\ref{sec:formalmodel}.
The 15 authentication and secrecy goals were each verified against the 40 models explained in Section~\ref{sec:compromise_scenarios}, resulting in 570 verification targets.
We will mostly discuss the authentication goals that failed in the partial compromise scenarios. While some of the failures are expected in the given scenarios, others provide new insights into the protocol and lead to recommendations for both implementations and future versions of the standard.


\definecolor{Gray}{gray}{0.9}
\newcolumntype{C}{>{\columncolor{Gray}}p{2.50em}}
\newcolumntype{S}{p{2.50em}}
\newcolumntype{R}{>{\raggedright\arraybackslash}p{17mm}}

\addtolength{\tabcolsep}{-3pt}
\begin{table}
\centering
\hypertarget{target:results_tables}{}

\begin{minipage}{1.0\textwidth}
\centering
\caption{Results for the default-server approach}
\label{tab:results-default_smdp}
{\footnotesize
\begin{tabular}{l|CSSCSCCSCCS|SCSC}
  \toprule
  \multicolumn{1}{l|}{\textbf{Partial compromise}} &
    \multicolumn{11}{c|}{\textbf{Authentication goals}} &
    \multicolumn{4}{c}{\textbf{Secrecy goals}} \\
    \hhline{~---------------}
  \multicolumn{1}{l|}{\textbf{scenario}} & A & B & B$'$ & C & D & E & F & G & I & J & K & W & X & Y & Z 
  \\
  \midrule
  1: --- & \bT{} & \bT{} & \bT{} & \bT{} & \bT{} & \bT{} & \bT{} & \bT{} & \bT{} & \bT{} & \bT{} & \bT{} & \bT{} & \bT{} & \bT{} \\
  2: server & \bF{\rr{2}} & \bF{\rr{2},\rr{c}} & \bT{} & \bF{\rr{2}} & \bF{\rr{c}} & \bF{\rr{2}} & \bF{\rr{2}} & \bF{2} & \bF{\rr{2}} & \bF{\rr{2}} & \bT{} & \bT{} & \bF{\rr{2}} & \bT{} & \bF{\rr{2}} \\
  3: eUICC & \bT{} & \bF{\rr{\rr{4}}} & \bT{} & \bO{\rr{d}} & \bF{\rr{4}} & \bT{} & \bT{} & \bF{\rr{4}} & \bT{} & \bT{} & \bT{} & \bF{\rr{4}} & \bT{} &\bF{\rr{4}} & \bT{} \\
  4: LPA & \bT{} & \bT{} & \bT{} & \bT{} & \bT{} & \bT{} & \bT{} & \bT{} & \bT{} & \bT{} & \bT{} & \bT{} & \bT{} & \bT{} & \bT{} \\
  5: 2nd server & \rO{\rr{3}} & \rO{\rr{c}} & \bT{} & \rO{\rr{3}} & \rO{\rr{c}} & \rO{\rr{3}} & \rO{\rr{3}} & \bT{} & \rO{\rr{3}} & \rO{\rr{3}} & \bT{} & \bT{} & \rO{\rr{3}} & \bT{} & \rO{\rr{3}} \\
  6: 2nd eUICC & \bT{} & \bT{} & \bT{} & \rO{\rr{d}} & \bT{} & \bT{} & \bT{} & \bT{} & \bT{} & \bT{} & \bT{} & \bT{} & \bT{} & \bT{} & \bT{} \\
  7: 2nd MNO & \bT{} & \bT{} & \bT{} & \bT{} & \bT{} & \bT{} & \bT{} & \bT{} & \bT{} & \bT{} & \bT{} & \bT{} & \bT{} & \bT{} & \bT{} \\
  8: order as user & \bT{} & \bT{} & \rF{\rr{7}} & \bT{} & \bT{} & \bT{} & \bT{} & \rF{\rr{7}} & \bT{} & \bT{} & \rF{\rr{7}} & \bT{} & \bT{} & \bT{} & \bT{} \\
  9: order for eUICC & \bT{} & \bT{} & \rF{\rr{a}} & \bT{} & \bT{} & \bT{} & \bT{} & \rF{\rr{a}} & \bT{} & \rF{\rr{a}} & \rF{\rr{a}} & \bT{} & \bT{} & \bT{} & \bT{} \\
  \bottomrule
  \multicolumn{16}{l}{Attacker owns some eUICCs in all the scenarios 1--9. Client-side goals are gray. No security is expected in Scenarios 2-3.}
\end{tabular}
}
\end{minipage}

\bigskip
\begin{minipage}{1.0\textwidth}
\centering
\caption{Results for the activation-code approach}
\label{tab:results-ac}
{\footnotesize
\begin{tabular}{l|CSSCSCCSCCS|SCSC}
  \toprule
  \multicolumn{1}{l|}{\textbf{Partial compromise}} &
    \multicolumn{11}{c|}{\textbf{Authentication goals}} &
    \multicolumn{4}{c}{\textbf{Secrecy goals}} \\
    \hhline{~---------------}
    \multicolumn{1}{l|}{\textbf{scenario}} & A & B & B$'$ & C & D & E & F & G & I & J & K & W & X & Y & Z 
  \\ \midrule
  1: --- & \bT{} & \bT{} & \rO{\rr{1}} & \bT{} & \bT{} & \bT{} & \bT{} & \rO{\rr{1}} & \bT{} & \bT{} & \rO{\rr{1}}  & \bT{} & \bT{} & \bT{} & \bT{} \\
  2: server & \bF{\rr{2}} & \bF{\rr{2},\rr{c}} & \bF{\rr{1},\rr{f}} & \bF{\rr{2}} & \bF{\rr{c}} & \bF{\rr{2}} & \bF{\rr{2}} & \bF{\rr{1},\rr{2},\rr{f}} & \bF{\rr{2}} & \bF{\rr{2}} & \bF{\rr{1},\rr{f}}  & \bT{} & \bF{\rr{2}} & \bT{} & \bF{\rr{2}} \\
  3: eUICC & \bT{} & \bF{\rr{4}} & \bF{\rr{1},\rr{6}} & \bO{\rr{d}} & \bF{\rr{4}} & \bO{\rr{e}} & \bO{\rr{e}} & \bF{\rr{1},\rr{4},\rr{6}} & \bO{\rr{e}} & \bO{\rr{e}} & \bF{\rr{1},\rr{6}}  & \bF{\rr{4}} & \bT{} & \bF{\rr{4}} & \bT{} \\
  4: LPA & \bT{} & \bT{} & \rF{\rr{1},\rr{9}} & \bT{} & \bT{} & \bT{} & \bT{} & \rF{\rr{1},\rr{9}} & \bT{} & \rF{\rr{9}} & \rF{\rr{1},\rr{9}}  & \bT{} & \bT{} & \bT{} & \bT{} \\
  5: 2nd server & \rO{\rr{3}} & \rO{\rr{c}} & \rO{\rr{1}} & \rO{\rr{3}} & \rO{\rr{c}} & \rO{\rr{3}} & \rO{\rr{3}} & \rO{\rr{1}} & \rO{\rr{3}} & \rO{\rr{3}} & \rO{\rr{1}}  & \bT{} & \rO{\rr{3}} & \bT{} & \rO{\rr{3}}\\
  6: 2nd eUICC & \bT{} & \rO{\rr{5}} & \rO{\rr{1}} & \rO{\rr{d}} & \rO{\rr{5}} & \bT{} & \bT{} & \rO{\rr{1},\rr{5}} & \bT{} & \bT{} & \rO{\rr{1}}  & \rO{\rr{5}} & \bT{} & \rO{\rr{5}} & \bT{} \\
  7: 2nd MNO & \bT{} & \bT{} & \rO{\rr{1}} & \bT{} & \bT{} & \bT{} & \bT{} & \rO{\rr{1}} & \bT{} & \bT{} & \rO{\rr{1}}  & \bT{} & \bT{} & \bT{} & \bT{}\\
  8: order as user & \bT{} & \bT{} &\rF{\rr{1},\rr{7}} & \bT{} & \bT{} & \bT{} & \bT{} & \rF{\rr{1},\rr{7}} & \bT{} & \bT{} & \rF{\rr{1},\rr{7}}  & \bT{} & \bT{} & \bT{} & \bT{} \\
  10: code leaks & \bT{} & \bT{} & \rF{\rr{1},\rr{8}} & \bT{} & \bT{} & \bT{} & \bT{} & \rF{\rr{1},\rr{8}} & \bT{} & \bT{} & \rF{\rr{1},\rr{8}}  & \bT{} & \bT{} & \bT{} & \bT{} \\
  11: code spoofed & \bT{} & \bT{} & \rF{\rr{1},\rr{b}} & \bT{} & \bT{} & \bT{} & \bT{} & \rF{\rr{1},\rr{b}} & \bT{} & \rF{\rr{b}} & \rF{\rr{1},\rr{b}}  & \bT{} & \bT{} & \bT{} & \bT{} \\
  \bottomrule
  \multicolumn{16}{l}{Attacker owns some eUICCs in all the scenarios 1--11.  Client-side goals are gray. No security is expected in Scenarios 2-3.}
\end{tabular}
}
\end{minipage}

\bigskip
\begin{minipage}{0.96\textwidth}
  \centering
  \caption{Summary of major vulnerabilities in RSP}
  \label{tab:results-summary}
  {\small
  \setlength{\extrarowheight}{2pt}
  \begin{tabularx}{0.95\textwidth}{Xp{2mm}RR}
    \toprule
    \textbf{Vulnerability} && \textbf{Attack references} & \textbf{Recommenda\-tions}
    \\ \midrule
    Security of RSP depends partly on the TLS tunnel. Compromise of TLS causes leak of the activation code and general failure in scenarios 5 and 6.
    &&
    {\scriptsize \rr{1},\rr{3},\rr{5},\rr{c},\rr{d},\rr{e}}
    &
    R10~(incl.~R1,~R2, R3,~R7,~R9)*
    \\
    The client (LPA and eUICC) may lack a-priori knowledge of the server OID. Communicating the OID out-of-band is optional in the activation-code approaches and not supported in the default-server approach.
    &&
    ${}^{\rr{3}}$
    &
    R1,R2,R7*
    \\
    The \SMDP server may lack a-priori knowledge of the eUICC identifier $U$. In that case, the security depends entirely on the secrecy of the activation code, which is prone to leaking and injection.
    &&
    ${}^{\rr{1},\rr{5},\rr{8},\rr{9},\rr{b},\rr{f}}$
    &
    R3*
    \\
    Misbinding vulnerabilities are created because several signed messages lack identity information on which the client and server should agree. Messages 4--7 should include the server OID. Messages 8--9 should include the eUICC identifier $U$.
    &&
    ${}^{\rr{c},\rr{d}}$
    &
    R7,R9*
    \\
    No reliable method for verifying user intent, including both the mobile subscriber identity and eUICC ownership. Many MNOs do not authenticate the user or verify the user's possession of the eUICC before ordering a SIM profile. Thus, the adversary could order a profile on behalf of another user or make an unauthorized order for their eUICC.
    &&
    ${}^{\rr{7},\rr{8},\rr{a},\rr{b}}$
    &
    R4,R5,R6
    \\\bottomrule
    *) \red{formally verified that recommendations together remove the vulnerability}
  \end{tabularx}
  }
\end{minipage}

\end{table}
\addtolength{\tabcolsep}{3pt}

Tables \ref{tab:results-default_smdp}--\ref{tab:results-ac} summarize the verification results for the two profile ordering and download initialization approaches explained in Section~\ref{sec:ordering_and_initialization}: the default-server approach and the activation-code approach. Each row in the tables corresponds to one of the compromise scenarios defined in Section~\ref{sec:compromise_scenarios}. The columns correspond to the security goals from Section~\ref{sec:formalgoals}. The columns for client-side goals are on gray background, while server-side goals are on white. The successfully verified security goals are marked with \bT{} or \bO{}, and the failed security goals are marked with \bF{}. \bO{} denotes goals that fail if TLS is disabled between the LPA and server. Red color highlights interesting failures.

As expected, \textit{all the security goals were successfully verified for the case where all the participants are honest and only the network is untrusted}, i.e., in scenario 1. A few goals in scenario 1 depend on TLS \resultstableref{\rr{1}}, and we will return to them in Section~\ref{sec:results_tls}. Otherwise, the rest of the discussion is about the partial compromise scenarios 2--11.

\subsection{Failures of the \SMDP server authentication}
\label{sec:results_server_authentication}

As can be expected, there is a catastrophic security failure in scenario 2, in which the server's private keys are compromised.
In the attack, the adversary plays the role of the server towards the eUICC, and the victim server does not need to be involved at all. This violates all the client-side correspondence properties \resultstableref{\rr{2}}. The profile secrecy goal (Sec Z) also fails because the client accepts a fake profile created by the adversary. \red{Moreover, when the victim server participates in the protocol, the attacker can cause inconsistent states between the server and eUICC by tampering with the signed messages from the server. This can cause the failure of the server-side correspondence properties Auth B and Auth G.}

The first somewhat surprising result is that the security of server authentication (Auth A) in scenario 5 depends on the TLS tunnel. Recall that the adversary in scenario 5 has compromised one \SMDP server, but not the one chosen by the MNO for the victim user. If TLS is not used, the adversary can redirect connections from the client to the compromised server. The consequence is that the client-side security properties fail \resultstableref{\rr{3}}, exactly as in scenario 2 \resultstableref{\rr{2}}. The client will even accept a fake profile from the compromised server. The relevant rows in Tables~\ref{tab:results-default_smdp}--\ref{tab:results-ac} show the similarity between scenarios 2 and 5 when TLS is not used. The failure in scenario 5 is more serious, however, because \textit{an adversary that has compromised an \SMDP server anywhere in the world can issue fake SIM profiles to any subscriber of any MNO} --- even if they have no business relation with the compromised server.

The root cause for the problem in scenario 5 is that, as we observed in Section~\ref{sec:AuthA}, the client typically lacks a-priori knowledge of the server OID. This allows the adversary to sign message 4--5 with any GSMA-issued server certificate. The RSP protocol has the option of delivering the server OID to the client together with the activation code so that the LPA can check it after receiving message 4. Unfortunately, there is currently no similar option for the default-server approach, where the eUICC stores only the server domain name.
We make the following recommendations:
\begin{itemize}
    \item[\recommend{R1}] \textit{In the activation-code approach, always provide the server OID to the LPA during the profile download initialization.}
    \item[\recommend{R2}] \textit{Amend the RSP specification for the default-server approach so that, in addition to the default server domain name $S$, either the LPA or the eUICC stores the default server OID and compares it with the value in the server certificate \cert{Sa} during the server authentication.}
\end{itemize}
The choice between implementing \recommend{R2} in the LPA or the eUICC depends on whether there is a secure path for configuring the correct server OID value into the eUICC. If the value can be configured securely in the device supply chain (together with the server domain name $S$), the eUICC should perform the check. On the other hand, if the correct server OID is input into the eUICC via the LPA, we might just as well trust the LPA to make the comparison with \cert{Sa}.

\subsection{Failures of client authentication and profile secrecy}
\label{sec:results_client_auth_and_profile_secrecy}

Auth B is a technical client authentication goal that covers only the authentication step.
As we might expect, the stepwise goal fails in scenario 3, where the eUICC's private key is compromised \resultstableref{\rr{4}}. This is because the eUICC authentication relies on the certificate and signature in message 6-7. The adversary in scenario 3 has the private key, and it can sign a spoofed message 7. In the activation-code approach, the adversary needs a valid activation code for the spoofed message; it can obtain one by ordering a profile for itself.

In scenario 6, the security of Auth B depends on TLS, but only in the activation-code approach \resultstableref{\rr{5}}. Recall that, in scenario 6, the adversary has compromised the private key of one eUICC, e.g., by reverse-engineering its own mobile device. Without TLS, the adversary can capture the activation code \Iac from an honest user. It then uses the captured code in a spoofed message 7, which it signs with the compromised eUICC's private key. This attack is possible in the activation-code approach because the server does not know the correct eUICC identifier $U$ and accepts any eUICC certificate \cert{U} in message 7.

While these failures of the client authentication (Auth B) are rather technical, they have serious consequences later in the protocol. Failure of the initial client Auth B leads to the failure of the client authentication in the key exchange (Auth D), which in turn breaks the secrecy of the session key (Sec W) and secrecy of the profile (Sec Y). In other words, the Diffie-Hellman session key is compromised and, therefore, \textit{the SIM profile leaks to the adversary}.

In order to avoid the failure in scenario 6, we make the following recommendation:
\begin{itemize}
  \item[\recommend{R3}] \textit{In the activation-code approach, always register the eUICC identifier $U$ to the server during the profile ordering.}
\end{itemize}

\subsection{Identity fraud and leaked activation code}
\label{sec:results_identity_fraud_and_leaked_code}

The high-level client-authentication goal Auth B$'$ compares the authentication results with the initial user intent and the order received by the server. The most obvious failure occurs again in scenario 3, where the client's private key has been compromised \resultstableref{\rr{6}}. The other failures of Auth B$'$, on the other hand, are quite different from the stepwise goal Auth B.

The other failures of Auth B$'$ are relatively easy to understand if we start the discussion from scenario 8, which models identity fraud in the profile ordering. In the attack \resultstableref{\rr{7}}, the adversary spoofs the victim user's identity when requesting a profile from the MNO. The adversary then downloads the profile to an adversary-owned eUICC. There is a slight technical difference between the default-server and activation-code approaches: in the former, the adversary makes the order explicitly for the adversary-owned eUICC, while in the latter, the adversary receives an activation code and uses it to download the profile into the adversary-owned eUICC. The end result of the attack is the same in both approaches.

Identity fraud in the profile ordering is not the only way in which the adversary can obtain an activation code that is associated with a victim user's account at the MNO. Scenario 10 models explicitly the leaks that may happen due to careless handling of the activation code by the victim user, MNO employees, or other parts of the supply chain. The consequent security failures \resultstableref{\rr{8}} are exactly the same as in the identity fraud case \resultstableref{\rr{7}}. Furthermore, in scenario 4, where the LPA software is compromised, the LPA can leak the code because delivering the activation code to the eUICC is one of the main features of the LPA. Again, the security failures are the same \resultstableref{\rr{9}}.

The practical consequences of the leaked activation code are serious, and we will highlight them here. With the activation code, the adversary can download the honest user's SIM profile into the adversary's mobile device. The adversary's mobile network access will then be charged to the victim user's account. Even more seriously, the adversary may be able to hijack the victim's mobile telephone number, answer and make calls at that number, and impersonate the victim in applications that use the telephone number for authentication. This type of attack is called \textit{SIM swapping}~\cite{BafoutsouENISA}. The general observation here is that identity fraud in the profile ordering phase can enable \textit{identity fraud} in other contexts where the mobile device is used for authentication.
Moreover, the attack is fairly realistic. When the order is made offline in a shop or over the phone, the primary objective of the MNO staff is to sell subscriptions and not to act as security officers. When the order is made online or with the LPA software, the strength of the user authentication depends on the supported identity verification methods, which depend on the MNO practices and national regulations.

We summarize the need for due diligence as a recommendation:
\begin{itemize}
  \item[\recommend{R4}] \textit{When the user requests a profile for their eUICC, the MNO should verify the user identity carefully. In the default-server approach, it should do this before ordering the profile from the \SMDP server, and in the activation-code approach, before giving the activation code to the user.}
\end{itemize}

The RSP specification does not explicitly state the above requirement because it is an MNO function and, thus, outside the technical scope of the specification. Nevertheless, it is essential for the security of the protocol. Additionally, \textit{the user, MNO, and the entire supply chain should treat the activation code as a critical secret}. This secrecy requirement for the activation code is already stated in the RSP specification.

\subsection{Unauthorized order for an eUICC}
\label{sec:results_unauthorized_order}

When a malicious user contacts the MNO to request a SIM profile for an eUICC, the user identity is not the only thing they can misrepresent. In the default-server approach, they may also lie about their eUICC identifier $U$. Scenario 9 models explicitly situations where the adversary requests a profile with its own user identity but for someone else's eUICC identifier. The client authentication goal (Auth B$'$) fails in this scenario \resultstableref{\rr{a}}, and so does the dependent goal for the full protocol (Auth K).

\begin{figure}[t]
  \centering
  \includegraphics[scale=0.9]{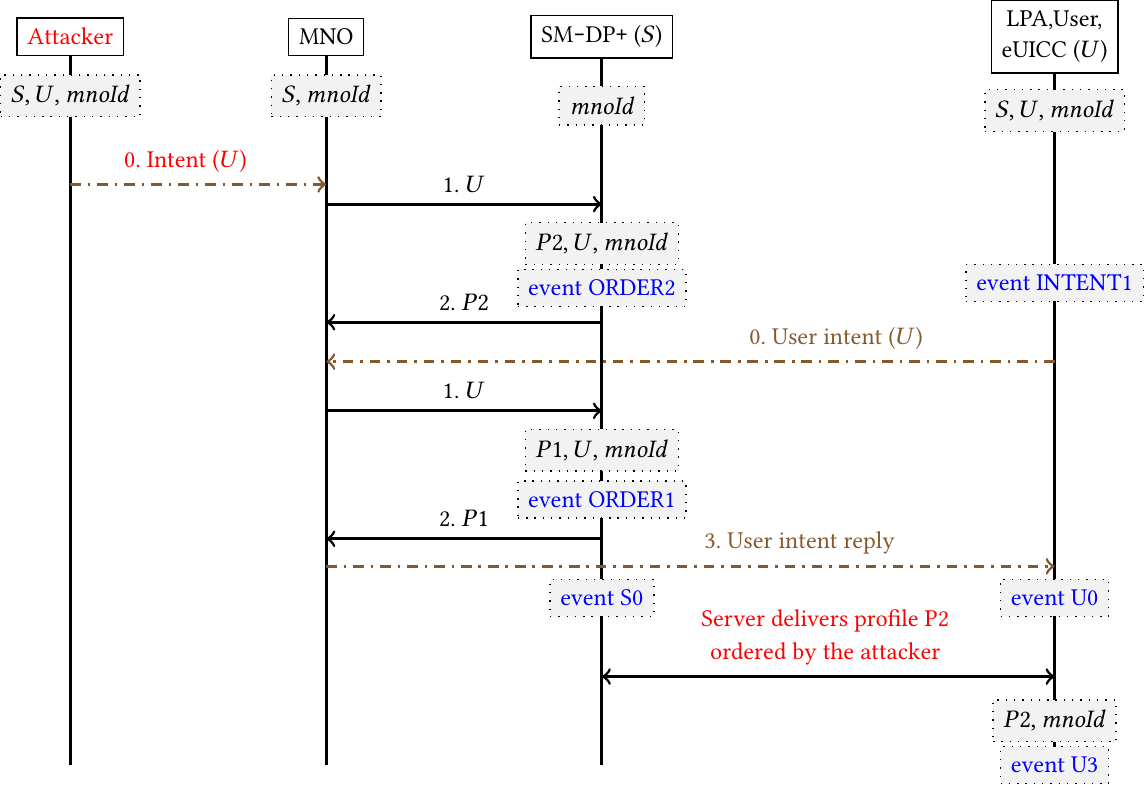}
  \caption{The adversary orders a second profile for the target eUICC. When the user initiates a download, the server may deliver the wrong profile to the eUICC. The attack works against the default-server approach to profile download initialization.}
  \Description{A user may wrongly order a profile from an MNO or an adversary may order a profile for an eUICC. These profiles may be served by the same server. When the user initiates a download, the server may deliver the wrong profile to the eUICC. The attack works against the default-server approach to profile download initialization.}
  \label{fig:fig-default-smdp-attack}
\end{figure}

The attack for the default-server approach is shown in Figure~\ref{fig:fig-default-smdp-attack}. The victim user orders a profile for their own eUICC. Meanwhile, the attacker has ordered a profile from the same MNO for its own user identity but for the victim's eUICC. There are now two profiles for the same eUICC on the same server, and it is unclear which profile the victim will download. Depending on the timing of the requests and the server implementation, the victim may end up installing the profile ordered by the attacker. The attack is possible if (i) the adversary can request a profile from an MNO without proving its possession of the eUICC, and (ii) the adversary knows the eUICC identifier $U$ of the target user. It seems likely that these conditions will be met in practical situations.

The practical consequences of the attack are quite troubling. The installed profile is associated with the adversary's mobile subscription, and thus the adversary gains some control over the victim's mobile access. It may be able to redirect the user's calls and text messages. It may also be able to access sensitive information about the victim's mobile access including call logs or location. Moreover, we should consider an extended scenario 9 where the user is careless about checking the \mnoid during the profile download (after message 12 in Figure~\ref{fig:commonhandshake}). In that case, the attacker can order its profile from a different MNO that uses the same server. This second MNO could be a compromised one that leaks session keys to enable call interception, or it could pay the adversary for bringing in new subscribers, albeit unwilling ones. These attacks violate some of the most fundamental security requirements of mobile networks: call confidentiality and billing integrity.

In the activation-code approach, similar failures happen in scenario 11, where the adversary can inject an activation code to the victim user \resultstableref{\rr{b}}. Moreover, the compromised LPA in scenario 4 could not only leak an activation code (see Section~\ref{sec:results_identity_fraud_and_leaked_code}) but also inject one \resultstableref{\rr{9}}. Again, the practical consequence is that the user may install a SIM profile that is associated with someone else's --- usually the adversary's --- mobile subscription.

To prevent the attacks, the MNO should verify that the user who requests the profile actually owns the eUICC with the specific identifier. Unfortunately, this verification is not easy to implement. In our formal model, the user is both the person who holds the mobile device containing the eUICC, and the mobile subscriber who pays be bills. In the real world, on the other hand, they can be two different entities. The intent to obtain a new SIM profile should, ideally, be authorized by both the subscriber and the person with the mobile device. There is currently no support for such dual authorization in RSP, and adding it would require major extensions to the protocol concepts, user interface, and messages.

While there is no strong solution, it is possible to mitigate the attacks. As the most important mitigation, \textit{users have to be careful not to use activation codes from untrusted sources.} Obviously, they should also not install LPA software from untrusted sources. Additionally, we recommend the following changes to the RSP specification:

\begin{itemize}
  \item[\recommend{R5}] \textit{Amend the RSP specification with design guidelines for a uniform user experience on how the user confirms the MNO before the profile download or installation.}
  \item[\recommend{R6}] \textit{Amend the RSP specification for the default-server approach so that, if there are multiple profiles on the \SMDP server, the user is warned and asked to choose the one to download. }
\end{itemize}

The RSP specification already supports the inclusion of the MNO icon and name in the profile metadata. The LPA or the UE may display the profile metadata to the user for confirmation. However, there is currently no guidance on how the information should be displayed, what the user should compare, or what the MNO icons and names look like.

From Tables~\ref{tab:results-default_smdp}--\ref{tab:results-ac}, we observe that scenarios 8, 9, 10, and 11 as well as scenario 4 in the activation-code approach have nearly the same failures. The explanation is that the server-side authentication goals require correspondence between initial user intent and the server's later session state for several parameters including the MNO, server, and eUICC identifiers. The attacks in the different scenarios cause a mismatch of different parameters, but the end result is always a mismatch between the user intent and the later server state, causing failure of the server-side goals Auth B$'$ and Auth K.

\subsection{Misbinding without TLS}
\label{sec:results_misbinding}

\begin{figure*}[t]
  \centering
  \includegraphics[scale=0.9]{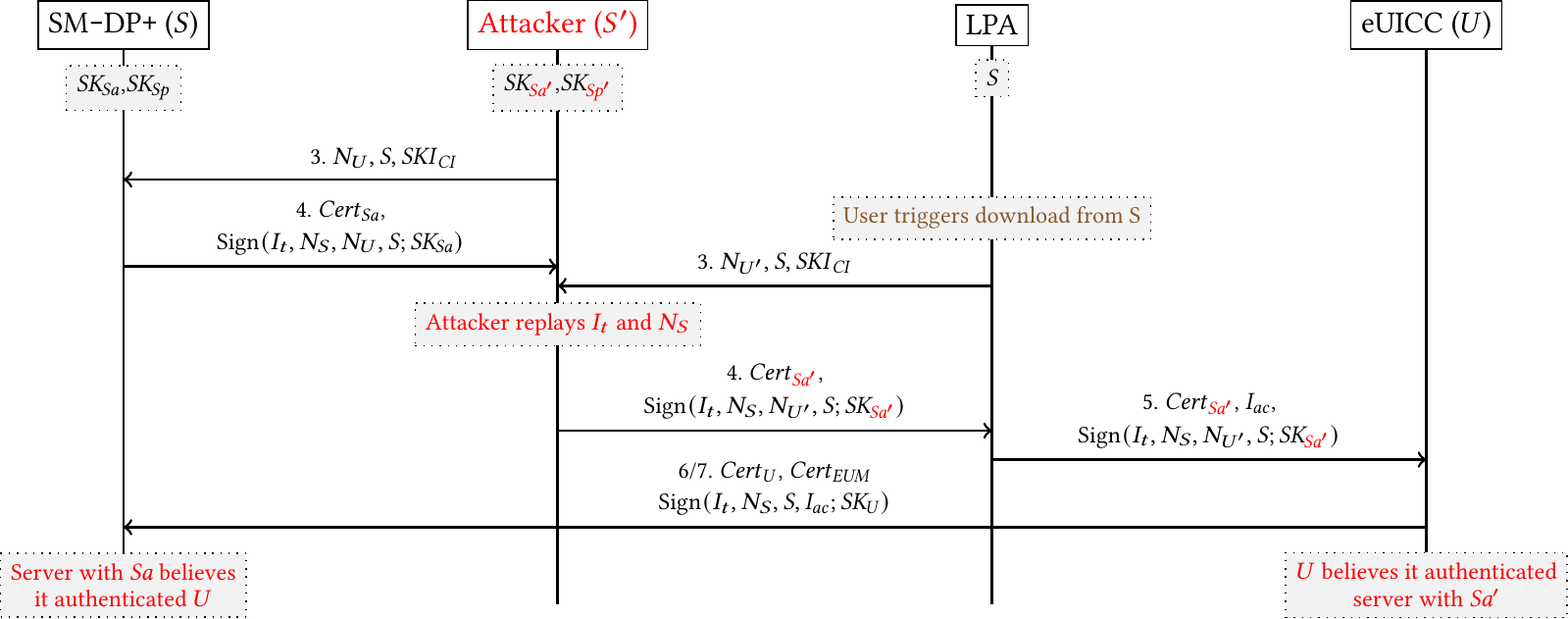}
  \caption{Misbinding attack in scenario 5 when TLS is not used. The adversary has compromised a second server $S'$. The eUICC and honest server end up with unmatching beliefs.}
  \Description{In scenario 5, the adversary has compromised an \SMDP server. In this attack, the adversary acts as a man in the middle between an eUICC and an honest server.
  The adversary can act as a server towards the eUICC because the eUICC has no prior knowledge about the server identity and it will accept any valid server certificate. The adversary starts a second session with the honest server and obtains a signed message 4. It replaces the signature of message 4 from the honest server with the signature of the compromised server. As a result, the honest eUICC believes it is talking with the compromised server, while the honest server thinks it is talking with the eUICC.}
  \label{fig:misbinding}
  \end{figure*}

In scenario 2, as well as in scenario 5 when TLS is not used, there is an additional failure of the stepwise client authentication (Auth B) due to a so-called \textit{misbinding} attack~\cite{krawczyk2003sigma, blake1999unknown, sethi2019misbinding} \resultstableref{\rr{c}}. In the attack, the eUICC is willing to perform mutual authentication with the authorized \SMDP servers $S$ and $S'$. The latter server is compromised, i.e., the adversary controls its private keys. The details of the misbinding attack are shown in Figure~\ref{fig:misbinding}. The result is that the eUICC thinks the session identified by $I_t$ is with the server whose private key is $Sa'$; on the other hand, the server whose private key is $Sa$ thinks it has the same session with the eUICC. This violates the correspondence property Auth B. Note that this is not normal server impersonation or man in the middle; in a sense, the adversary gets the honest server $S$ to impersonate the compromised server $S'$.

The misbinding attack causes inconsistent states between the legitimate protocol participants. The practical consequence could be confusion among the users and MNO service staff. In this case, the failure of Auth B leads to the further failure of Auth D. The consequences are mitigated by the fact that misbinding attacks, in general, do not cause leakage of secrets.

To prevent the misbinding, the protocol messages should be more explicit about the server identity. We recommend the following:

\begin{itemize}
  \item[\recommend{R7}] \textit{Amend the RSP specification so that the signed messages in the initial authentication (messages 4--5 and 6--7) include the server OID, which the message recipient compares with the expected value. This requires the adoption of recommendations} \recommend{R1} \textit{and} \recommend{R2}.
\end{itemize}

\red{As mentioned in Section~\ref{sec:handshake}, the RSP specification does not require the server to check value $S$ in message 7. We verified models with and without the check and did not find any impact on the security goals. Nevertheless, it seems prudent to perform the comparison:}

\begin{itemize}
  \item[\recommend{R8}] \textit{\red{Amend the RSP specification so that the server compares $S$ in the received message 7 with its domain name.}}
\end{itemize}

\begin{figure}[t]
  \centering
  \includegraphics[scale=0.9]{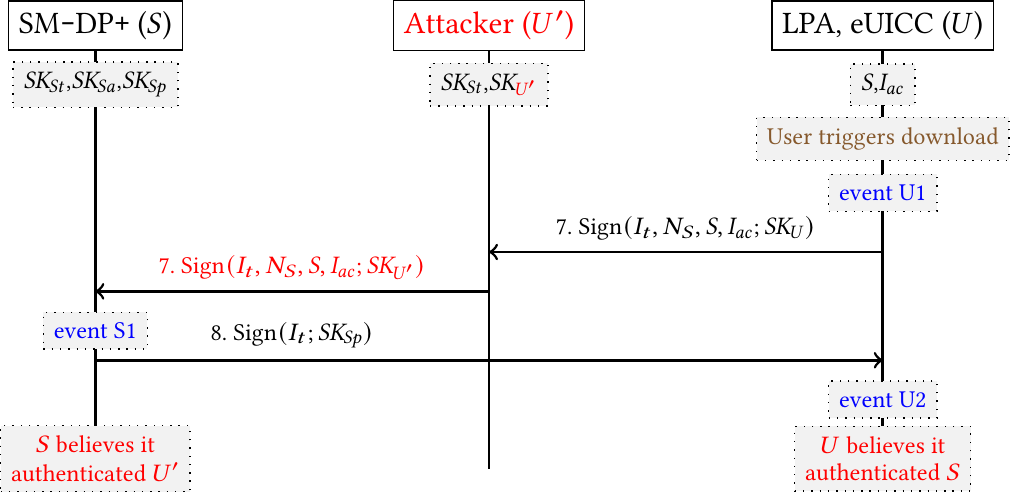}
  \caption{Misbinding attack in scenario 6 when TLS is not used. The adversary replaces the eUICC signature on message 7 with one by a compromised second eUICC. The eUICC and server end up with unmatching beliefs. Similar attacks are also present in scenario 3}
  \label{fig-attack-compromised-server-tls-key-auth-c}
  \Description{In scenario 6, the adversary has compromised an eUICC. In this attack, the adversary acts as a MitM between an honest eUICC and a server. The adversary can act as a client towards the server as the server has no prior knowledge about the honest eUICC identity and it will accept any valid eUICC certificate. The adversary replaces the signature of message 7 from the honest eUICC with the signature of the compromised eUICC. As a result, the honest client believes it is talking with the server, while the server thinks it is talking with the compromised eUICC.}
\end{figure}

There is also a misbinding vulnerability in the reverse direction. Without the TLS tunnel, Auth C fails in scenarios 3 and 6 \resultstableref{\rr{d}}, in which the adversary has compromised the private key of at least one eUICC. We first explain the attack in the activation-code approach, as shown in Figure~\ref{fig-attack-compromised-server-tls-key-auth-c}. The adversary intercepts message 7 from the honest eUICC $U$. It replaces the signature and certificate on message 7 with those of a compromised eUICC~$U'$ and forwards the modified message to the server. When the server receives message 7, it believes the identity information in the eUICC certificate. The server responds with message 8, which only contains the session identifier~$I_t$, from which $U$ cannot tell that the message was intended for~$U'$. The result is that $U$ thinks the session identified by~$I_t$ is with~$S$, while~$S$ thinks the session is with the $U'$, which violates the correspondence property Auth C. In the default-server approach, the adversary will first prepare for the attack by ordering a profile for the compromised eUICC $U'$ to the same server~$S$.

The consequences of the second misbinding are limited because the attack will be detected during the following key exchange, where the server checks that message 11 is signed by the same eUICC as the earlier message 7. Nevertheless, we recommend increasing the robustness of the profile binding step:

\begin{itemize}
  \item[\recommend{R9}] \textit{Amend the RSP specification so that the server includes the eUICC identifier $U$ from the initial authentication (message 7) in the signed profile-binding (message 8--9), and that the eUICC checks the identifier upon receipt of the message.}
\end{itemize}

\subsection{Dependence on the TLS tunnel}
\label{sec:results_tls}

We have already seen some failures that occur if the TLS tunnel is disabled \resultstableref{\rr{3},\rr{5},\rr{c},\rr{d}}. Let us now return to the pure network adversary (scenario 1). In the default-server approach, the security goals do not depend on the TLS tunnel. In the activation-code approach, on the other hand, \textit{TLS is needed to protect the activation code}. If TLS is not used, the activation code leaks, which results in the previously mentioned failures \resultstableref{\rr{1}} with the same consequences as in scenario 10 \resultstableref{\rr{8}}. Naturally, any failure of scenario 1 occurs in all the other scenarios as well; this can be seen in Table~\ref{tab:results-ac} columns B$'$, G, and K where the security always fails without TLS. The root causes of these failures are (i) that the server does not initially know the eUICC identity~$U$ and relies on the secret activation code to authorize the eUICC, and (ii) that without TLS, the adversary can capture the activation code and use it to download the profile to a different eUICC.

In addition to stealing the activation code, the adversary can replace the code with the wrong one. In scenario 3, where the adversary has the private key of the honest user's eUICC, this causes some additional failures of client-side security properties \resultstableref{\rr{e}}.
In the attack, the adversary intercepts message 7, replaces the activation code \Iac in it, signs the modified message, and forwards it to the server.

It is important to realize that the model without the TLS tunnel is also a model for situations where the security of the TLS tunnel is compromised. As explained in Section~\ref{sec:compromise_scenarios}, the TLS server and its private key are relatively vulnerable to hacking. The effect of the compromised tunnel can be seen in scenarios 2 and 4 as well, where the server and client end of the tunnel are compromised, respectively. The failures in these scenarios with TLS \resultstableref{\rr{f},\rr{9}} are similar to the failures in scenario 1 without TLS \resultstableref{\rr{1}}.

RSP is an unusual cryptographic protocol in that its security depends partly on the encapsulation of the messages into a TLS tunnel. Looking at the symbols \bO{} in Tables~\ref{tab:results-default_smdp}-\ref{tab:results-ac}, we see that quite a few vulnerabilities arise in the absence of the TLS tunnel between the LPA and \SMDP server. The protocol would be more robust if it was designed to be independent of the TLS transport. We recommend the following:
\begin{itemize}
  \item[\recommend{R10}] \textit{Redesign the RSP protocol so that it does not depend on the TLS channel for any critical security requirements. This can achieved by implementing recommendations} \recommend{R1}, \recommend{R2}, \recommend{R3}, \recommend{R7} \textit{and} \recommend{R9}.
\end{itemize}

More precisely, in the default-server approach, implementing recommendations \recommend{R2}, \recommend{R7}, and \recommend{R9} removes the dependence on TLS. \red{In a variant of the protocol model that implements these recommendations, the verification results} for the default-server approach change so that there is no difference between using or not using TLS (i.e., \bO{} becomes \bT{} in Table~\ref{tab:results-default_smdp}). \red{We modeled \recommend{R2} so that the LPA stores the default server OID.}

In the activation-code approach, implementing recommendations \recommend{R1}, \recommend{R3}, \recommend{R7}, and \recommend{R9} similarly removes the dependence on TLS (except in scenario 3), \red{which we also verified.} The reader might not expect such a big improvement in the activation-code approach because TLS protects the confidentiality of the activation code. The explanation is that recommendation \recommend{R3} removes the client-authentication role of the activation code and thus the need to keep it secret; instead, the client authentication will depend on the pre-registered eUICC identifier $U$ at the server and on the eUICC certificate \cert{U} in messages 6-7. (In scenario 3, \recommend{R3} is ineffective due to the compromised eUICC private key. This should have no impact on the protocol design decisions because scenario 3 is irreparably insecure due to the other attacks covered in Section~\ref{sec:results_client_auth_and_profile_secrecy}.)

\red{To further investigate the role of TLS, we experimented with a variant of the protocol where TLS is replaced by the server signing message 4 with its private TLS certificate \cert{St}. The idea is to create a two-directional binding between the identities $Sa$ and $S$. When combined with recommendations \recommend{R8} and \recommend{R9}, this protocol modification achieved the same as \recommend{R10} in the default server approach. In a sense, this shows that TLS is not needed in the default server approach except to bind the server domain name S to its identifier Sa.}

Notwithstanding the argument made above, the TLS provides an additional layer of encryption that protects user privacy. Without the tunnel, RSP would leak the eUICC identifier and certificate to the communication network. TLS also contributes towards the broad requirement for confidentiality protection that was quoted in Section~\ref{sec:AuthD}. Therefore, we suggest keeping the tunnel but relegating it to only a privacy protection layer.

\subsection{Failures of install notification}
\label{sec:results_notification}

The install notification takes place after the activation of the SIM profile, and any security failures at this stage are not catastrophic. If the server accepts a spoofed notification, it and the MNO may wrongly believe that the profile was installed but not in use. The result could be a discrepancy between the MNO's business information systems and the mobile network state. It is unlikely that the user's access to the network would be affected. We summarize the attacks for completeness.

Authentication of the profile install notification (Auth G) can fail for several different reasons. First, it depends on the TLS tunnel. Second, attacks against the client authentication goals Auth B or B' can be continued against Auth G \resultstableref{\rr{1},\rr{4},\rr{5},\rr{6},\rr{7},\rr{8},\rr{9},\rr{a},\rr{b},\rr{f}}. The only exception is that the misbinding attack against Auth B \resultstableref{\rr{c}} does not propagate to Auth G because the profile install notification message 15, unlike message 7 (Figure~\ref{fig:commonhandshake}), contains the server OID of the intended recipient.

\section{Discussion}
\label{sec:discussion}

This section summarizes the vulnerabilities identified in our research. It also discusses the importance of partial compromise scenarios, the limitations of the modeling and verification methods, and the possibility of simplifying the RSP protocol.

\subsection{Summary of the main vulnerabilities}
\label{sec:discussion_vulnerabilities}

 We will now discuss the significance of the most important vulnerabilities discovered in Section~\ref{sec:results}. They are also summarized in Table~\ref{tab:results-summary}.

\paragraph{Dependence on TLS}
Our most important observation is that the security of RSP depends on the TLS tunnel from the mobile device to the \SMDP server and that this dependence is mostly unnecessary. In Section~\ref{sec:results_tls}, we recommend a clear path to removing this dependence. In the default-server approach, there is hardly any reason not to implement the recommended changes (\recommend{R2}, \recommend{R7}, and \recommend{R9}). In the activation-code approach, the recommended changes (\recommend{R1}, \recommend{R3}, \recommend{R7}, and \recommend{R9}) would similarly make RSP independent of TLS, but the decision to implement these changes is not as straightforward as in the default-server approach. The convenience of activation codes depends on the fact that they are distributed to potential customers without knowing in advance who will install the profile and on which device. This advantage would be lost with \recommend{R3}, which requires pre-registration of the eUICC identifier for each activation code. An alternative way to remove the dependence on TLS is to encrypt the activation code (message 7 in Figure~\ref{fig:commonhandshake}) on the RSP level. This is possible but will require restructuring of the cryptographic protocol so that the Diffie-Hellman key exchange and server authentication take place before the client sends the activation code to the server.

\paragraph{No a-priori knowledge of identifiers}
The next two vulnerabilities in Table~\ref{tab:results-summary} are related to the lack of a-priori knowledge about the peer identity. There is some overlap with the TLS dependence discussed above, but the design consideration is nevertheless different. The client and server in RSP may not initially know each other's identifiers. When that is the case, they have no way of knowing whether they are talking to the intended entity. The authentication with certificates does assure the server that it is talking with some eUICC, and it does assure the client that it is talking with some \SMDP server in the world. The weakness is that they cannot differentiate between the intended entity and other entities of the same class that may be operated by the adversary.

A general solution to this problem is pre-established identities, which can then be authenticated with the certificates. The server OID certainly should be pre-distributed to the LPA or eUICC (recommendations \recommend{R1} and \recommend{R2}). As already discussed above, pre-registering the eUICC identifier (\recommend{R3}) is not as obvious because it would require major changes to the business processes in the activation-code approach. Without \recommend{R3}, the entire path of the activation code in RSP must protect its confidentiality. This path includes the \SMDP server \resultstableref{\rr{f}}, the MNO sales and support staff or online channels and the mobile user \resultstableref{\rr{8},\rr{b}}, the LPA \resultstableref{\rr{9}}, the eUICC \resultstableref{\rr{5}}, as well as the TLS tunnel from LPA back to the server \resultstableref{\rr{1}} --- or redesigning the protocol to protect the confidentiality without TLS, as already suggested above.

\paragraph{Misbinding vulnerability}
The next vulnerability in Table~\ref{tab:results-summary} is a simple protocol design flaw that enables the misbinding attacks. The solution is to include the server and eUICC identities in the messages (recommendations \recommend{R7} and \recommend{R9}).

\paragraph{Difficulty of verifying user intent}
The final vulnerability in Table~\ref{tab:results-summary} is more fundamental. RSP relies on the user deciding which profile they want to install. For the protocol designer, this creates multiple difficult problems which were discussed in Sections~\ref{sec:results_identity_fraud_and_leaked_code} and~\ref{sec:results_unauthorized_order}.

The first problem is that the MNO needs to verify the identity and authorization of the user who requests a profile from it. The communication with the user may take place offline, or online without strong authentication. Moreover, the MNO and its employees are incentivized to make the profile ordering process as easy as possible, and they may prefer getting new business to performing identity checks.

A further problem is that the user intent should be authorized both by the mobile subscriber and by the person holding the mobile device with the eUICC. These are not always the same person or entity. The difficulty of verifying both the subscriber and eUICC owner intent reliably may be one of the reasons why the default-server approach is currently not in wide use. \textit{An advantage of the activation-code approach is that the MNO gives the code to the subscriber and it is entered into the LPA by the person holding the mobile device, which implicitly ensures the authorization from both user roles.}

Finally, the users themselves could make mistakes, such as leaking the activation code or accepting the wrong code. There is a step in the protocol where the user should check the MNO identifier before the profile download. This gives the user another opportunity to stop a download that does not match the user's intent, but a typical mobile user may not be able to check the MNO identifier reliably.


\subsection{Do failures in the partial compromise scenarios matter?}
\label{sec:discussion_realistic}

In scenarios 2-3, the private keys of one of the protocol endpoints are compromised. The serious failures of client authentication \resultstableref{\rr{6}} as well as server authentication and profile secrecy \resultstableref{\rr{2},\rr{4}} are quite expected, and it would be unrealistic to defend against such strong attackers.

In scenario 3, we can compare RSP with the analogous attacks against the old, removable SIM cards. In scenario 3, the adversary has compromised the physical integrity or supply chain of the target eUICC and extracted its private key. As a consequence, the adversary can capture the SIM profile including the shared secret in the profile. This is comparable to the adversary compromising the integrity of an old-fashioned SIM card or its supply chain and extracting the pre-shared secret in it. Thus, we can argue that RSP does not create a new vulnerability in scenario 3.

The other partial compromise scenarios 4-10, on the other hand, are more realistic, and the security failures in them should be taken seriously.

Scenario 5 assumes that there is one compromised \SMDP server somewhere in the world, which seems quite possible. This leads to failure of the server authentication \resultstableref{\rr{3},\rr{c}}. Scenario 6, where the adversary has compromised the private keys of one or more eUICCs somewhere in the world, is almost guaranteed to occur. The resulting attacks break not only the client authentication but also the secrecy of the authentic SIM profile \resultstableref{\rr{5},\rr{d}}. The security failures in these scenarios are prevented by running RSP inside the TLS tunnel, but as we have argued, that is a relatively weak protection, and the protocol should not depend on TLS for security.

In scenario 6, we can again compare RSP with removable SIM cards. In scenario 6, \textit{the adversary has extracted the private key of one eUICC anywhere in the world, and this enables it to capture the SIM profile of any target eUICC with a network attack}. There is no similar weakness in the old removable SIM cards. The attack is possible against MNOs that distribute activation codes without associating the code with a specific eUICC, which is a common practice.

One historical concern in mobile networks has been SIM cloning~\cite{singh2013gsm,simcloning,zhou2013need}. Modern removable SIM cards are resistant to the published cloning methods~\cite{singh2013gsm}. The closest equivalent attack against RSP is the leaked profile in scenario 3 or 6 \resultstableref{4,5}. The attacker could then continue by emulating an eUICC with the stolen profile or by injecting the profile into another eUICC in scenario 2 or 5 \resultstableref{2,3}.

Scenarios 8-10 seem quite likely to occur in practice because they model mistakes by the human participants including the mobile user and MNO employees. Compared with the removable SIM card, the attacks \resultstableref{\rr{7},\rr{8},\rr{a},\rr{b}} are similar to stealing the victim's SIM card and replacing the victim's SIM with the wrong one. Even when the SIM profile is successfully protected by the secure hardware module, which can be either the eUICC or the removable SIM card, the module may end up in the wrong user's mobile device. The new issue in RSP is that the SIM profile is a digital object rather than a physical one, which means that the current methods for identifying the object and authorizing actions need to be replaced with digital ones.

\subsection{Limitations of the modeling and verification}
\label{sec:discussion_limitations}

A formal protocol model is necessarily an abstraction of the full specification and leaves out many details. It is essential to identify the protocol participants and the important messages between them.

The RSP protocol has various participants: the eUICC, LPA, user, MNO, and server. On one hand, we can model these entities as separate processes and can consider separately the compromise of each entity and each communication channel between them. On the other hand, by combining multiple such entities into one process, the model becomes easier to understand and more manageable for the verification tools, but some subtle attacks might go undetected. We have modeled the LPA and user as one combined process.

The protocol messages in the model must be similarly optimized. The first challenge was to identify the critical messages in RSP and their parameters from the information scattered all over the specification. Even then, it was necessary to optimize the model so that only the essential messages are included. As one major optimization, we do not model the negotiation of the cryptographic algorithms (and thus might miss downgrade attacks~\cite{bhargavan2016downgrade}) or error handling. \red{Similarly, we did not model the client privacy goals. The RSP specification requires the client not to reveal any private data to an unauthenticated server. Verifying this goal fully would require a more detailed model and analysis of the message data fields and message flows in error situations.}

\red{
There were a few cases where ProVerif was unable to complete the verification without additional help. First, in some cases, ProVerif found a security goal violation but was unable to construct an attack trace, and in others, it produced a trace but was unable to prove the violation. In all such cases, ProVerif output was sufficient for us to understand and confirm the security failure. Second, commutativity in ECDH was modeled in the standard way as an equation, and equations can cause problems for the verifier. In scenarios 3 and 6, the attacker knows an eUICC private key \SK{U} and can thus send a false public ECDH parameter $Q_U$ in message 11. In these scenarios, ProVerif was unable to complete the verification of goals Auth E, Auth F, and Auth G. This was resolved by restricting the model so that any public ECDH parameter $Q_S$ sent by a server in message 12 cannot be replayed as $Q_U$ in message 11. 
\begin{equation*}
   \texttt{restriction } \event{SENT}{Q_S} \AND \event{RECEIVED}{Q_U} \IMPLY Q_S<>Q_U
\end{equation*}
The events RECEIVED and SENT were added to the server after receiving $Q_U$ in message 11 and before sending $Q_S$ in message 12, respectively. We argue that this restriction to the model does not mask any attacks because the attacker can achieve just as much by sending a random value as $Q_U$ as by replaying a $Q_S$ from a parallel session or another server. Either way, the eUICC will reject message 12--13 and the attacker cannot construct a valid one because it does not know the session keys $k, k'$. In practice, it is always more beneficial for the attacker to use a $Q_U$ for which it knows the private parameter $d_U$ because then it can compute the session keys and potentially continue the attack.
}

Sometimes, the specification does not fully define the participant's behavior, and we had to make assumptions about the implementation. For example, the LPA software acts as an intermediary between the server and eUICC. As shown in Figure~\ref{fig:commonhandshake}, it checks the value of $S$ in message 4 and \mnoid in message 12 before passing them on to the eUICC. The RSP specification does not say whether LPA should perform any other sanity checks on the messages, such as verifying the signatures and comparing other message fields with their expected values. We experimented with both a relaxed LPA that checks only these two values and otherwise passes through any messages without looking at their contents, and a strict LPA that checks all the values it can. The verification results did not indicate any difference in the security of the two LPA models, but the strict model produced more verification results.

The TLS channel between the LPA and the server was particularly difficult to model accurately. First, only the server is authenticated while the LPA is anonymous. Second, we might expect the tunnel to protect the integrity of the session inside, but since the communication is implemented as an HTTP API, TLS will actually protect only the individual request-response pairs. Any continuity of state between the requests has to be implemented by the application-layer protocol, in this case, RSP. We experimented with different models of the TLS tunnel. Fortunately, we did not find cases where subtle differences in the TLS model would impact the verification results.

Furthermore, the RSP protocol has alternative modes and optional features that cause variation in the message flows and content. We solve this problem by modeling a large number of protocol variants (the default-server and activation-code approaches and the optional features relevant to the recommendations). Additional model variants were created by the partial compromise scenarios. To maintain consistency between the model variants, we generate all the models from one combined source. A theoretical limitation that arises from this way of modeling the protocol variants is that we will not detect potential adverse interactions between different protocol variants.

In addition to the protocol participants and messages, the security goals were also formalized. The queries were optimized iteratively, on one hand, to detect any attacks, and on the other hand, to verify the security properties that can be verified. The interesting queries are often ones that fail in one scenario and succeed in another.
One difficulty is that the first security failure found by the verification tool may be a trivial one. In that case, it is necessary to experiment with queries that ignore the first discovered attack, so that more serious or more profound failures can be uncovered. The queries shown in this paper have many possible variations, e.g., on which parameters to include in the injective correspondences. The queries included in Section~\ref{sec:formalgoals} are the ones that best demonstrate the analysis results of Section~\ref{sec:results}; many further variants were evaluated during our research work.

In general, the modeling process involves iterative inclusion and exclusion of details from the protocol model and security goals as well as optimization of their representation.

\subsection{SM-DS assisted profile ordering and download initialization}
\label{sec:security_SM_DS}

As mentioned in Section~\ref{sec:ordering_and_initialization}, the RSP specification includes a third, \textit{SM-DS assisted} approach to the profile ordering and download initialization. We do not model this approach because it has not been widely deployed as an open system. Rather, individual vendors have deployed proprietary variants of this approach.

The SM-DS assisted approach relies on an additional hierarchical entity, the subscriber manager discovery servers (SM-DS), to assist the mobile device in finding the profiles that are waiting for it in the \SMDP servers. The LPA receives one or more \textit{event identifiers} from the SM-DS servers. The rest of the protocol is similar to the activation-code approach with a pre-registered eUICC identifier (recommendation \recommend{R3}). If the SM-DS and \SMDP server are closely coupled as a proprietary solution, the security would correspond to this combination. In a more open model, we would have to model also the partial compromise scenario in which some of the SM-DS servers are compromised.

\subsection{Simplifying the RSP protocol}
\label{sec:simplified_protocol}

The common handshake and profile download, as seen in Figure~\ref{fig:commonhandshake}, comprise more credentials, messages, and roundtrips than a typical authentication protocol.

The multiple different server certificates (see Table~\ref{tab:certificate_notation}) in the consumer RSP protocol have their origin in the machine-to-machine RSP specification~\cite{RSP-02}. In M2M RSP, two technically and commercially separate entities, the subscription manager secure routing (SM-SR) and the subscription manager data preparation (SM-DP), perform the entity authentication and session key agreement, respectively. In contrast, consumer RSP combines these roles into a single commercial entity, the \SMDP. Thus, the separate key pairs and steps for entity authentication and profile binding are not strictly necessary, and the profile binding would not need to be a separate step in the protocol.

The key agreement and authenticated encryption (messages 10-13) in the RSP protocols borrow from the GlobalPlatform defined SCP11a~\cite{SCP11} specification. This explains why they are a separate phase rather than merged with entity authentication.

Thus, the consumer RSP protocol could be simplified. For example, we could combine the profile binding (messages 8-9) with the server authentication (messages 4-5) and the client key-exchange messages (10-11) with the client authentication (messages 6-7). The dependence on TLS could be removed completely by encrypting the handshake messages similar to TLS 1.3~\cite{rescorla2018transport}. However, such major changes to the protocol design will require a thorough consultation with the stakeholders in the standards process. We believe that the recommendations made in this paper provide a much shorter path to increasing the robustness of RSP security.

\section{Conclusion}
\label{sec:conclusion}

We have modeled and analyzed the GSMA consumer remote SIM provisioning protocol that allows remote downloading of SIM profiles to eUICCs. Our work includes constructing a formal model of the RSP protocol based on its specifications, identifying assumptions and implicit security goals in the specification, formalizing the security goals, and defining realistic partial compromise scenarios where the protocol should be tested. We performed verification of 15 selected security goals with the ProVerif tool against two protocol variants in 11 scenarios, all with and without a TLS tunnel. In addition to presenting the verification results and describing potential attacks against the protocol, we discussed the root causes of the failures and made practical recommendations for improving the robustness of the RSP protocol specification and its implementations. The results show the power of formal modeling and verification in clarifying the goals of security protocols and in understanding the assumptions made in their design. 

\Urlmuskip=0mu plus 1mu\relax
\apptocmd{\sloppy}{\hbadness 10000\relax}{}{}
\bibliographystyle{ACM-Reference-Format}
\bibliography{paper}


\appendix
\section{Appendix (to be removed)}
\label{sec:Appendix4}

\begin{figure*}
 \centering
 \includegraphics[scale=0.9]{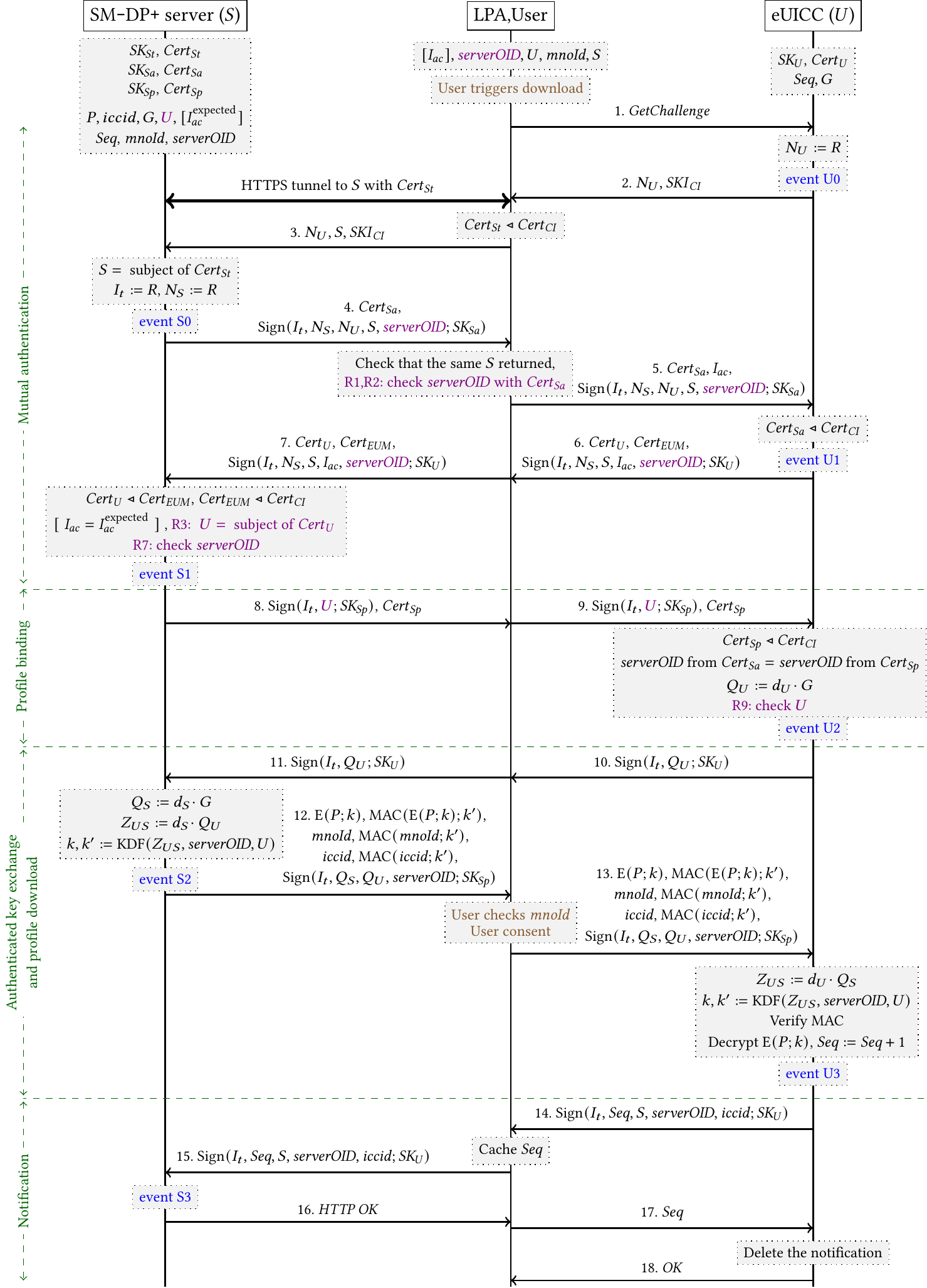}
 \caption{Common handshake and profile download in the RSP protocol with recommendation R10 (changes in \Rec{violet})}
 \Description{Message sequence chart of the main part of the RSP protocol with enabled recommendations. The communication endpoints are server and the eUICC. The communication passes though the LPA software on the mobile device.}
\label{fig:commonhandshakewithrecommendations}
\end{figure*}

\end{document}